\documentclass[prd,nofootinbib,reprint,preprintnumbers,showkeys]{revtex4-1}
\usepackage{xcolor}
\usepackage{amsmath,amsfonts,bm}
\usepackage{graphicx, epstopdf,scrextend}
\usepackage{mathrsfs,mathtools}

\definecolor{darkgreen}{rgb}{0,0.4,0} 
\definecolor{darkblue}{rgb}{0,0,0.6} 
\usepackage[colorlinks=true,linkcolor=darkblue,citecolor=darkgreen]{hyperref}

\usepackage{pgfplots,pgfplotstable}
\usepgfplotslibrary{fillbetween}
\usetikzlibrary{intersections}
\pgfplotsset{compat=newest}

\pgfplotsset{every axis/.style={
    width=12cm,
    height=10cm,
    grid=both,
    scaled ticks=false,
    yticklabel style={/pgf/number format/.cd, fixed,precision=5}
  }
}

\newcommand{\as}{\alpha_{\mathrm{s}}}

\newcommand{\LA}{\mathrm{A}}
\newcommand{\LB}{\mathrm{B}}
\newcommand{\LF}{\mathrm{F}}

\newcommand{\scH}{\textsc{h}}
\newcommand{\LI}{\mathrm{I}}

\newcommand{\LR}{\mathrm{R}}

\newcommand{\LT}{\mathrm{T}}
\newcommand{\La}{\mathrm{a}}
\newcommand{\Lb}{\mathrm{b}}
\newcommand{\Lc}{\mathrm{c}}
\newcommand{\Ld}{\mathrm{d}}
\newcommand{\Lf}{\mathrm{f}}
\newcommand{\Lg}{\mathrm{g}}
\newcommand{\Lh}{\mathrm{h}}

\newcommand{\Lp}{\mathrm{p}}
\newcommand{\Ls}{\mathrm{s}}
\newcommand{\Lu}{\mathrm{u}}

\newcommand{\LZ}{\mathrm{Z}}

\newcommand{\MSbar}{\overline{\mathrm{MS}}}

\newcommand{\GeV}{\ \mathrm{GeV}}
\newcommand{\TeV}{\ \mathrm{TeV}}

\newcommand{\cF}{\mathcal{F}}

\newcommand{\cH}{\mathcal{H}}

\newcommand{\cN}{\mathcal{N}}
\newcommand{\cO}{\mathcal{O}}
\newcommand{\cP}{\mathcal{P}}

\newcommand{\cS}{\mathcal{S}}

\newcommand{\cU}{\mathcal{U}}
\newcommand{\cV}{\mathcal{V}}

\definecolor{red}{rgb}{1,0,0}
\newcommand{\NOTE}[1]{\textcolor{red}{ \bf[NOTE: #1]}}

\def\mi{{\mathrm i}}

\def\ket#1{\big|{#1}\big\rangle}
\def\bra#1{\big\langle{#1}\big|}
\def\<>#1{\big\langle{#1}\big\rangle}
\def\[]#1{\big[{#1}\big]}

\def\sket#1{\big|{#1}\big)}
\def\sbra#1{\big({#1}\big|}
\def\sbrax#1{\big({#1}}        

\newbox\charbox
\newbox\slabox
\def\s#1{{ 
        \setbox\charbox=\hbox{$#1$}
        \setbox\slabox=\hbox{$/$}
        \dimen\charbox=\ht\slabox
        \advance\dimen\charbox by -\dp\slabox
        \advance\dimen\charbox by -\ht\charbox
        \advance\dimen\charbox by \dp\charbox
        \divide\dimen\charbox by 2
        \raise-\dimen\charbox\hbox to \wd\charbox{\hss/\hss}
        \llap{$#1$}
}}

\usepackage{pgfplots,pgfplotstable}

\newcommand{\ratioplot}[8][]{
  \pgfplotstableread{#3}\datatable

  \addplot[name path=pluserror,draw=none,no markers,forget plot]  table
  [ x={#4},
    y expr=\thisrow{#5}/\thisrow{#7} + (\thisrow{#6}/abs(\thisrow{#7})+\thisrow{#8}*abs(\thisrow{#5})/\thisrow{#7}**2)
  ]{\datatable};

  \addplot[name path=minuserror,draw=none,no markers,forget plot]  table
  [ x={#4},
    y expr=\thisrow{#5}/\thisrow{#7} - (\thisrow{#6}/abs(\thisrow{#7})+\thisrow{#8}*abs(\thisrow{#5})/\thisrow{#7}**2)
  ]{\datatable};

  \addplot[forget plot,#2] fill between[on layer={},of=pluserror and minuserror];
  \addplot [#1,no markers] table [x={#4},y expr=\thisrow{#5}/\thisrow{#7}]{\datatable};
}

\pgfplotsset{every axis/.style={
    width=8.6cm,
    height=8.6cm,
    grid=both,
    scaled ticks=false,
    yticklabel style={/pgf/number format/.cd, fixed,precision=5}
  }
}

\newif\ifusefigs
\usefigstrue

\begin{document}

\title{Evolution of parton showers and parton distribution functions}

\author{Zolt\'an Nagy}

\affiliation{
  DESY,
  Notkestrasse 85,
  22607 Hamburg, Germany
}

\email{Zoltan.Nagy@desy.de}

\author{Davison E.\ Soper}

\affiliation{
Institute for Fundamental Science,
University of Oregon,
Eugene, OR  97403-5203, USA
}

\email{soper@uoregon.edu}

\begin{abstract}
Initial state evolution in parton shower event generators involves parton distribution functions. We examine the probability for the system to evolve from a higher scale to a lower scale without an initial state splitting. A simple argument suggests that this probability, when multiplied by the ratio of the parton distributions at the two scales, should be independent of the parton distribution functions. We call this the {\em PDF property}. We examine whether the PDF property actually holds using \textsc{Pythia} and \textsc{Deductor}. We also test a related property for the \textsc{Deductor} shower and discuss the physics behind the results.
\end{abstract}

\date{22 July 2020}

\preprint{DESY-20-015}

\maketitle

\section{Introduction}
\label{sec:intro}

Parton shower algorithms for simulation of hadron-hadron collisions use ``backward evolution'' for the initial state part of the shower \cite{sjostrand}. The parton distribution functions (PDFs) for the initial state hadrons appear in the initial state splitting probabilities in the shower. In the simplest approximation, there is a property that relates two functions: (1) the probability for the state to evolve from a hard scale $\mu^2_\Lh$ to a softer scale $\mu^2_\Ls$ without an initial state splitting  and (2) the ratio of the PDFs at the two scales. We will call this property (defined below) the {\em PDF property}. 

Both PDF evolution and initial state parton shower evolution can be viewed as evolution under scale changes and both are calculated from perturbative splitting of initial state partons. Thus these two sorts of evolution are connected.  The PDF property represents a self-consistency condition for the shower in the sense that (as we shall see) if the property does not hold then there is some physics missing from the shower evolution. In this paper, we investigate this hypothesized property using the parton shower event generator \textsc{Pythia} \cite{pythia}. Then we investigate the same property using our own shower event generator \textsc{Deductor} \cite{Deductor}.

We state what the PDF property is in Sec.~\ref{sec:PDFproperty} and provide some alternative PDF sets for the purpose of testing the PDF property in Sec.~\ref{sec:pdfs}. Then we carry out the test using \textsc{Pythia} in Sec.~\ref{sec:pythiatest} and using \textsc{Deductor} with $k_\LT$ ordering in Sec.~\ref{sec:deductortest}. We test another property that applies to cross sections including threshold effects in \textsc{Deductor} in Sec.~\ref{sec:threshold}. In Sec.~\ref{sec:discussion}, we discuss the physics behind the results that we have seen. In Sec.~\ref{sec:otherordering}, we use \textsc{Deductor} to examine the previous properties using the default, virtuality based, shower ordering variable of \textsc{Deductor} and using angle ordering. Finally, we present a brief summary in Sec.~\ref{sec:conclusions}. We outline details of the operators used in \textsc{Deductor} in Appendix \ref{sec:operators}.

\section{The PDF property}
\label{sec:PDFproperty}

The property that we discuss is stated and derived in the widely used text of 
Ellis, Stirling, and Webber \cite{pinkbook}. To keep the notation uncomplicated, we consider a simplified case in which there is only one kind of parton and there is only one hadron. 

The parton distribution functions obey the Dokshitzer-Gribov-Lipatov-Altarelli-Parisi (DGLAP) evolution equation
\begin{equation}
\begin{split}
\label{eq:DGLAP}
&\mu^2 \frac{d}{d\mu^2}\, f(x, \mu^2)
\\ &\quad
= \int_0^1\!dz\ \frac{\as(\mu^2)}{2\pi}\, \widehat P(z) 
\left[
\frac{1}{z}\,f(x/z,\mu^2) - f(x,\mu^2)
\right]
.
\end{split}
\end{equation}
Here $\widehat P$ is the unregulated evolution kernel. We have applied a + prescription, which is reflected in the subtraction at $z = 1$.

Let us consider a parton shower algorithm based on a measure $\mu^2$ of the hardness of parton splittings. For instance, $\mu^2$ is often chosen to be a transverse momentum variable $k_\LT^2$. The scale $\mu^2$ is used to order splittings within the shower. Harder splittings come first, then softer splittings.

In the parton shower, we define a probability not to split between two scales by
\begin{equation}
\begin{split}
\label{eq:simplePi}
&\Pi(\mu_{\Ls}^2, \mu_{\Lh}^2, x)
\\& =
\exp\!\left(
- \int_{\mu_{\Ls}^2}^{\mu_{\Lh}^2}\!\frac{d\mu^2}{\mu^2}
\int\!\frac{dz}{z}\, \frac{\as(\mu^2)}{2\pi} \widehat P(z)
\frac{f(x/z, \mu^2)}{f(x,\mu^2)}
\right)
.
\end{split}
\end{equation}
Here $x$ is the momentum fraction of the initial state parton at the starting scale $\mu_\Lh^2$. Note the appearance of a ratio of PDFs in the exponent \cite{sjostrand}. There should be limits, $z_-(\mu^2) < z < z_+(\mu^2)$ for the $z$ integration, but we ignore that here. We simply treat $z_-(\mu^2)$ as being so close to 0 and $z_+(\mu^2)$ as being so close to 1 that the limits do not matter. We also define
\begin{equation}
\begin{split}
\label{eq:simplePipert}
\Pi_\textrm{pert}&(\mu_{\Ls}^2, \mu_{\Lh}^2, x)
\\ &=
\exp\!\left(
- \int_{\mu_{\Ls}^2}^{\mu_{\Lh}^2}\!\frac{d\mu^2}{\mu^2}
\int\!dz\,  \frac{\as(\mu^2)}{2\pi}\, \widehat P(z)
\right)
.
\end{split}
\end{equation}
Here we have the perturbative splitting function but no PDFs.

We can relate these functions by using the evolution equation for the PDFs. We have
\begin{equation}
\begin{split}
\label{eq:simplePiderivation}
\frac{f(x, \mu_\Ls^2)}{f(x, \mu_\Lh^2)} ={}& 
\exp\!\left(
- \int_{\mu_\Ls^2}^{\mu_\Lh^2}\!d\mu^2\
\frac{d \log[f(x, \mu^2)]}{d\mu^2}
\right)
\\
={}& \exp\!\Bigg(
- \int_{\mu_\Ls^2}^{\mu_\Lh^2}\! \frac{d\mu^2}{\mu^2}
\int\! dz\ \frac{\as(\mu^2)}{2\pi}
\\&\quad\quad \times
\widehat P(z)
\left[
\frac{f(x/z,\mu^2)}{z f(x, \mu^2)} - 1
\right]
\Bigg)
\\
={}&
\frac{\Pi(\mu_\Ls^2, \mu_\Lh^2, x)}
{\Pi_\textrm{pert}(\mu_\Ls^2, \mu_\Lh^2, x)}
\;.
\end{split}
\end{equation}
This tells us that if we define $\Pi_\textrm{alt}(\mu_\Ls^2, \mu_\Lh^2, x)$ by
\begin{equation}
\label{eq:Pialt}
\Pi_\textrm{alt}(\mu_\Ls^2, \mu_\Lh^2, x)
=
\frac{f(x, \mu_\Lh^2)}{f(x, \mu_\Ls^2)}\,
\Pi(\mu_\Ls^2, \mu_\Lh^2, x)
\;,
\end{equation}
then we will have 
\begin{equation}
\label{eq:PialtPipert}
\Pi_\textrm{alt}(\mu_\Ls^2, \mu_\Lh^2, x) 
= \Pi_\textrm{pert}(\mu_\Ls^2, \mu_\Lh^2, x)
\;.
\end{equation}
Then $\Pi_\textrm{alt}(\mu_\Ls^2, \mu_\Lh^2, x)$ will be independent of which PDF set is used in its calculation.

This is the PDF property in the simplified case. Now suppose we have two hadrons in the initial state and we have more than one kind of parton. We initiate a parton shower with two partons that create a hard scattering at scale $\mu_\Lh^2$. The two partons have momentum fractions $x_\La$, $x_\Lb$ and flavors $a$, $b$. We define $\Pi(\mu_\Ls^2, \mu_\Lh^2; x_\La,a,x_\Lb,b)$ to be the probability that, according to the parton shower algorithm used, this state evolves to a lower scale $\mu_\Ls^2$ with no parton splittings. 

We now define an alternative probability function by
\begin{equation}
\begin{split}
\label{eq:Pialtdef}
&\Pi_\textrm{alt}(\mu_\Ls^2, \mu_\Lh^2; x_\La,a,x_\Lb,b)
\\&\  =
\frac{f_{a/A}(x_\La, \mu_\Lh^2)\,f_{b/B}(x_\Lb, \mu_\Lh^2)}
{f_{a/A}(x_\La, \mu_\Ls^2)\,f_{b/B}(x_\Lb, \mu_\Ls^2)}\,
\Pi(\mu_\Ls^2, \mu_\Lh^2; x_\La,a,x_\Lb,b)
\;.
\end{split}
\end{equation}
The PDF property that we seek to investigate states that $\Pi_\textrm{alt}(\mu_\Ls^2, \mu_\Lh^2; x_\La,a,x_\Lb,b)$ is independent of the PDF set that we use in the calculation.\footnote{We can also demand that $\Pi_\textrm{alt}(\mu_\Ls^2, \mu_\Lh^2; x_\La,a,x_\Lb,b)$ depend on the parton momenta $p_\La = x_\La p_\LA$ and $p_\Lb = x_\Lb p_\LB$ but not on the hadron momenta $p_\LA$ and $p_\LB$. This amounts to demanding that $\Pi_\textrm{alt}$ be unchanged if we use new PDFs with rescaled momentum fractions: $\tilde f_{a/A}(x_\La,\mu^2) = \lambda_\LA f_{a/A}(\lambda_\LA x_\La,\mu^2)$ and $\tilde f_{b/B}(x_\Lb,\mu^2) = \lambda_\LB f_{b/B}(\lambda_\LB x_\Lb,\mu^2)$ for constants $\lambda_\LA$ and $\lambda_\LB$. Thus this is a special case of the demand that $\Pi_\textrm{alt}$ be independent of the PDFs.}

The text by Campbell, Huston, and Krauss \cite{blackbook} (Section 5.3.1.5) also relates the evolution of PDFs, Eq.~(\ref{eq:DGLAP}), to the no-splitting function in shower evolution, Eq.~(\ref{eq:simplePi}), using essentially the argument given above, but does not relate these to Eqs.~(\ref{eq:Pialt}) and (\ref{eq:PialtPipert}), so that we do not encounter the PDF property directly.

The plausibility argument for the PDF property given above assumes that the PDFs evolve according to the appropriate first order evolution kernel, which matches, at least approximately, the parton splitting functions used in a first order parton shower. We would like to check numerically whether there are any differences in $\Pi_\textrm{alt}$ values obtained with different PDF choices that are large enough to be proportional to $\as$ times possible logarithms. If one were to use PDFs that obey an evolution equation including order $\as^2$ contributions to the evolution kernel, then the difference of $\Pi_\textrm{alt}$ values calculated with two different PDF choices would certainly get contributions proportional to $\as^2$ times logarithms. These contributions would presumably be numerically rather small. However, we do not want such contributions to be present at all, so in this paper we use just first order evolution for the PDFs except when we use the standard CT14 NLO set \cite{CT14} in \textsc{Pythia}, where NLO denotes next-to-leading order. (We note that, in order to be consistent between shower evolution and PDF evolution, our program, \textsc{Deductor}, uses first order PDF evolution inside the shower.)

\section{Choices for parton distribution functions}
\label{sec:pdfs}

In order to test the PDF property, we need some choices for PDFs. The choices that we use do not need to fit data, but they do need to obey the standard DGLAP evolution equations. 

As a standard choice, we use the CT14 PDFs \cite{CT14}. These {\em do} fit data. We define alternative ``hard'' and ``soft'' PDF sets by setting
\begin{equation}
\begin{split}
\label{eq:hardpdf}
f^\mathrm{hard}_{\bar \Lu/\Lp}(x,\mu_\LI^2) ={}& \frac{1}{8}\,\frac{(1-x)^2}{x}
\;,
\\
f^\mathrm{hard}_{\Lu/\Lp}(x,\mu_\LI^2) ={}& f_{\bar \Lu/\Lp}(x,\mu_\LI^2)
+ 6\,(1-x)^2
\;,
\\
f^\mathrm{hard}_{\bar \Ld/\Lp}(x,\mu_\LI^2) ={}& \frac{1}{8}\,\frac{(1-x)^2}{x}
\;,
\\
f^\mathrm{hard}_{\Ld/\Lp}(x,\mu_\LI^2) ={}& f_{\bar d/\Lp}(x,\mu_\LI^2)
+ 3\,(1-x)^2
\;,
\\
f^\mathrm{hard}_{\Lg/\Lp}(x,\mu_\LI^2) ={}& \frac{1}{4}\,\frac{(1-x)^2}{x}
\;,
\end{split}
\end{equation}
and
\begin{equation}
\begin{split}
\label{eq:softpdf}
f^\mathrm{soft}_{\bar \Lu/\Lp}(x,\mu_\LI^2) ={}& \frac{35}{48}\,\frac{(1-x)^6}{x}
\;,
\\
f^\mathrm{soft}_{\Lu/\Lp}(x,\mu_\LI^2) ={}& f_{\bar u/\Lp}(x,\mu_\LI^2)
+ 14\,(1-x)^6
\;,
\\
f^\mathrm{soft}_{\bar \Ld/\Lp}(x,\mu_\LI^2) ={}& \frac{35}{48}\,\frac{(1-x)^6}{x}
\;,
\\
f^\mathrm{soft}_{\Ld/\Lp}(x,\mu_\LI^2) ={}& f_{\bar d/\Lp}(x,\mu_\LI^2)
+ 7\,(1-x)^6
\;,
\\
f^\mathrm{soft}_{\Lg/\Lp}(x,\mu_\LI^2) ={}& \frac{35}{24}\,\frac{(1-x)^6}{x}
\end{split}
\end{equation}
at an initial scale $\mu_\LI = 1.295 \GeV$. We define $f^\mathrm{hard}_{a/\Lp}(x,\mu_\LI^2) = f^\mathrm{soft}_{a/\Lp}(x,\mu_\LI^2) = 0$ for other flavors of partons. These functions obey the standard momentum and flavor sum rules. The first set is designated ``hard'' because the functions approach zero rather slowly as $x \to 1$, while the second set is designated ``soft'' because the functions approach zero rather quickly as $x \to 1$. We define $f^\mathrm{hard}_{a/\Lp}(x,\mu^2)$  and $f^\mathrm{soft}_{a/\Lp}(x,\mu^2)$ for $\mu^2 > \mu_\LI^2$ by solving the first order DGLAP equation. We insert standard flavor thresholds as in the CT14 set and use a rescaled argument of $\as$ according to the prescription in Eq.~(\ref{eq:CMW}) below. We use $\as(M_\LZ^2) = 0.118$.

\section{Test of the PDF property using Pythia}
\label{sec:pythiatest}

In this section, we test the PDF property using \textsc{Pythia} \cite{pythia} (version~8.2.43). The hardness parameter for \textsc{Pythia} is $\mu^2 = k_\LT^2$ with a particular definition of the transverse momentum $k_\LT^2$ in a splitting.  In \textsc{Pythia}, we set $\as(M_\LZ^2) = 0.118$ to match the coupling used in the parton evolution. 

We create a hard scattering event for proton-proton collisions at $\sqrt s = 13 \TeV$ corresponding to the Drell-Yan process with off-shell $Z_0$ and $\gamma$ production with an $e^+$$e^-$ final state. The starting scale $\mu_\scH^2$ for the shower is the square of the $e^+$$e^-$ mass. We accept events with $e^+$$e^-$ mass in the range $2.5\TeV < \mu_\scH < 2.7\TeV$. The parton momentum fractions $x_\La$ and $x_\Lb$ are determined by $Q^0 + Q^3$ and $Q^0 - Q^3$, where $Q$ is the $e^+$$e^-$ momentum. We have $\mu_\scH^2 = x_\La x_\Lb s$, so that $\sqrt{x_\La x_\Lb} \approx 0.2$. We accept events with $2/3 < x_\La/x_\Lb < 3/2$. The parton flavors can be $(a,b) = (\mathrm{u}, \bar{\mathrm{u}})$, $(\bar{\mathrm{u}}, \mathrm{u})$, $(\mathrm{d}, \bar{\mathrm{d}})$, $(\bar{\mathrm{d}}, \mathrm{d})$, or sometimes other choices. For each event, we record $x_\La$, $a$, $x_\Lb$, and $b$. Let $P(x_\La,a,x_\Lb,b)$ denote the probability density that the parton variables take the indicated values.

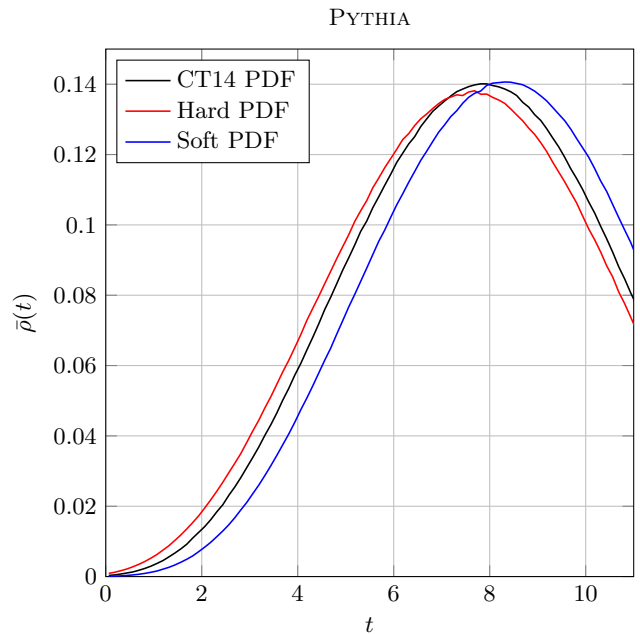
\begin{figure}
\begin{center}

\ifusefigs 

\begin{tikzpicture}
  \begin{axis}[
    title=\textsc{Pythia},
    xlabel={$t$}, ylabel={$\bar\rho(t)$},
    legend cell align=left,
    every axis legend/.append style = {
      at={(0.02,0.98)},
      anchor=north west
    },
    xmin=0, xmax=11.0,
    ymin=0, ymax=0.15
    ]
    
\pgfplotstableread{
 0.0625	0.00032315
0.1875	0.0004718075
0.3125	0.0007253775
0.4375	0.001012925
0.5625	0.001312925
0.6875	0.0017723
0.8125	0.002307125
0.9375	0.0029389250000000002
1.0625	0.0037041
1.1875	0.004516475
1.3125	0.0055644250000000004
1.4375	0.0066678
1.5625	0.007851525
1.6875	0.00919265
1.8125	0.01096375
1.9375	0.0125845
2.0625	0.01422
2.1875	0.0161695
2.3125	0.018390249999999997
2.4375	0.02040175
2.5625	0.02307
2.6875	0.025454
2.8125	0.028108
2.9375	0.030991499999999998
3.0625	0.033857750000000006
3.1875	0.036813250000000006
3.3125	0.040094000000000005
3.4375	0.043128
3.5625	0.046465
3.6875	0.05014925
3.8125	0.053490750000000004
3.9375	0.057243249999999996
4.0625	0.060561500000000004
4.1875	0.0643475
4.3125	0.06800375
4.4375	0.07184375
4.5625	0.07594224999999999
4.6875	0.07945975
4.8125	0.08326825
4.9375	0.08715575
5.0625	0.0906435
5.1875	0.09428725
5.3125	0.098174
5.4375	0.10108
5.5625	0.104855
5.6875	0.1079975
5.8125	0.11126
5.9375	0.11440249999999999
6.0625	0.11776249999999999
6.1875	0.12038499999999999
6.3125	0.123015
6.4375	0.12511
6.5625	0.1279525
6.6875	0.1301425
6.8125	0.1322925
6.9375	0.1338
7.0625	0.13525749999999997
7.1875	0.136635
7.3125	0.1378775
7.4375	0.13862
7.5625	0.13932250000000002
7.6875	0.13981
7.8125	0.140105
7.9375	0.1400975
8.0625	0.13962249999999998
8.1875	0.13925500000000002
8.3125	0.1386475
8.4375	0.137865
8.5625	0.136785
8.6875	0.13486
8.8125	0.1335
8.9375	0.13133
9.0625	0.12971749999999999
9.1875	0.12734
9.3125	0.1247325
9.4375	0.122195
9.5625	0.11923249999999999
9.6875	0.115995
9.8125	0.11315750000000001
9.9375	0.110045
10.062	0.106545
10.188	0.10334750000000001
10.312	0.09960525
10.438	0.096023
10.562	0.09225725
10.688	0.0881035
10.812	0.08462975
10.938	0.080654
11.062	0.07713199999999999
11.188	0.07319225
11.312	0.06930475
11.438	0.0652735
11.562	0.06166225
11.688	0.058109499999999994
11.812	0.05445
11.938	0.0506455
}\PythiaDYtdistCT

\pgfplotstableread{
 0.0625	0.000899165
0.1875	0.001229875
0.3125	0.0016595
0.4375	0.0021449750000000004
0.5625	0.0027725249999999996
0.6875	0.0034661249999999996
0.8125	0.004275950000000001
0.9375	0.00521865
1.0625	0.006239975
1.1875	0.007477175000000001
1.3125	0.008743850000000001
1.4375	0.010246
1.5625	0.011811499999999999
1.6875	0.01353925
1.8125	0.015366
1.9375	0.01729775
2.0625	0.019507
2.1875	0.021823250000000002
2.3125	0.02423125
2.4375	0.026783
2.5625	0.02952125
2.6875	0.032190750000000004
2.8125	0.034943
2.9375	0.03813325
3.0625	0.041372000000000006
3.1875	0.044425
3.3125	0.047881
3.4375	0.05108725
3.5625	0.054632
3.6875	0.0581925
3.8125	0.06147325
3.9375	0.06508575
4.0625	0.06875375
4.1875	0.07277425
4.3125	0.07625475
4.4375	0.0796805
4.5625	0.0832835
4.6875	0.08682999999999999
4.8125	0.090415
4.9375	0.09381475
5.0625	0.09719325000000001
5.1875	0.10096250000000001
5.3125	0.10404
5.4375	0.1069
5.5625	0.110515
5.6875	0.11332249999999999
5.8125	0.11626500000000001
5.9375	0.1188675
6.0625	0.12148250000000001
6.1875	0.1242675
6.3125	0.12596749999999998
6.4375	0.128255
6.5625	0.13015749999999998
6.6875	0.131545
6.8125	0.13308999999999999
6.9375	0.13449250000000001
7.0625	0.1356725
7.1875	0.136345
7.3125	0.1369775
7.4375	0.13679
7.5625	0.137805
7.6875	0.1381375
7.8125	0.137155
7.9375	0.1371975
8.0625	0.1362525
8.1875	0.135515
8.3125	0.134565
8.4375	0.1329775
8.5625	0.13127250000000001
8.6875	0.1295925
8.8125	0.127525
8.9375	0.12571
9.0625	0.12339499999999999
9.1875	0.1208725
9.3125	0.1178375
9.4375	0.115035
9.5625	0.112245
9.6875	0.10918749999999999
9.8125	0.1061975
9.9375	0.1024975
10.062	0.09888649999999999
10.188	0.0957375
10.312	0.09228425
10.438	0.0883495
10.562	0.08521125
10.688	0.08093225
10.812	0.0776035
10.938	0.073753
11.062	0.06992174999999999
11.188	0.06642975000000001
11.312	0.06271449999999999
11.438	0.059048500000000004
11.562	0.055702
11.688	0.052123499999999996
11.812	0.048933500000000005
11.938	0.04549725
}\PythiaDYtdistHard

\pgfplotstableread{
0.0625	0.00007529675
0.1875	0.00012847
0.3125	0.0001968075
0.4375	0.0003131425
0.5625	0.00046748250000000003
0.6875	0.0006496625
0.8125	0.0008967825
0.9375	0.0012119
1.0625	0.00156445
1.1875	0.0020844
1.3125	0.00262025
1.4375	0.0032797
1.5625	0.0040621
1.6875	0.004918
1.8125	0.005886525
1.9375	0.007097175
2.0625	0.00837155
2.1875	0.009745575
2.3125	0.011333
2.4375	0.0129415
2.5625	0.014711499999999999
2.6875	0.0167115
2.8125	0.01893
2.9375	0.02107275
3.0625	0.02354725
3.1875	0.025981749999999998
3.3125	0.0286785
3.4375	0.03147775
3.5625	0.03442025
3.6875	0.037474
3.8125	0.040534
3.9375	0.04384625
4.0625	0.0474
4.1875	0.050696
4.3125	0.054205750000000004
4.4375	0.0578685
4.5625	0.06152275
4.6875	0.064968
4.8125	0.06893350000000001
4.9375	0.07276325
5.0625	0.07666675
5.1875	0.0802315
5.3125	0.0839685
5.4375	0.08770325000000001
5.5625	0.09151000000000001
5.6875	0.095316
5.8125	0.09853475
5.9375	0.10217
6.0625	0.10569500000000001
6.1875	0.10883999999999999
6.3125	0.112125
6.4375	0.1152025
6.5625	0.1183275
6.6875	0.120965
6.8125	0.12370249999999999
6.9375	0.126385
7.0625	0.1284925
7.1875	0.1307775
7.3125	0.132455
7.4375	0.1345925
7.5625	0.1362925
7.6875	0.13743999999999998
7.8125	0.1380625
7.9375	0.13966499999999998
8.0625	0.140305
8.1875	0.1404625
8.3125	0.1406325
8.4375	0.1405775
8.5625	0.140215
8.6875	0.139855
8.8125	0.1390225
8.9375	0.138025
9.0625	0.1367025
9.1875	0.13526
9.3125	0.133395
9.4375	0.1315075
9.5625	0.12963750000000002
9.6875	0.12726
9.8125	0.12451999999999999
9.9375	0.12189
10.062	0.1192625
10.188	0.1159875
10.312	0.11248
10.438	0.109495
10.562	0.10580999999999999
10.688	0.1020975
10.812	0.098549
10.938	0.09501475000000001
11.062	0.090815
11.188	0.08714125
11.312	0.08316024999999999
11.438	0.07943175
11.562	0.07530675
11.688	0.071367
11.812	0.067646
11.938	0.06357375
}\PythiaDYtdistSoft

    \addplot[black,semithick] table {\PythiaDYtdistCT};
    \addlegendentry{CT14 PDF}
    \addplot[red,semithick] table {\PythiaDYtdistHard};
    \addlegendentry{Hard PDF}
    \addplot[blue,semithick] table {\PythiaDYtdistSoft};
    \addlegendentry{Soft PDF}

  \end{axis}
\end{tikzpicture}

\fi

\end{center}
\caption{
Probability distribution of first splittings, $\bar\rho(t)$, versus shower time according to \textsc{Pythia}. Here and in the following figures, we use three PDF sets.
}
\label{fig:rho}
\end{figure}

We use the ``user hooks'' mechanism of \textsc{Pythia} to determine the scale $\mu_\Ls^2$ of the first initial state splitting. It is convenient to define a corresponding logarithmic variable
\begin{equation}
t = \log(\mu_\scH^2/\mu_\Ls^2)
\end{equation}
for the ``shower time'' of the first splitting. Thus $t = 0$ corresponds to a splitting at the scale of the hard interaction and larger $t$ corresponds to a softer first splitting. Let $\rho(t;x_\La,a,x_\Lb,b)dt$ denote the probability that, for the indicated choice of parton variables, the first splitting happens between $t$ and $t + dt$. Then the averaged probability that the first splitting happens between $t$ and $t + dt$ is $\bar\rho(t) dt$ where
\begin{equation}
\bar\rho(t) = \sum_{a,b}\int\!dx_\La \int\!dx_\Lb\
\rho(t;x_\La,a,x_\Lb,b)\, P(x_\La,a,x_\Lb,b)
\;.
\end{equation}
Then it is of interest to plot $\bar\rho(t)$ versus $t$. The plots for each of our choices of PDF sets are shown in Fig.~\ref{fig:rho}. We would expect that with harder PDFs, it is more likely to have a first splitting at a smaller value of $t$. That is what we see.

\begin{figure}
\begin{center}

\ifusefigs 

\begin{tikzpicture}
  \begin{axis}[
    title=\textsc{Pythia},
    xlabel={$t$}, ylabel={$\overline \Pi(t)$},
    legend cell align=left,
    every axis legend/.append style = {
      anchor=north east
    },
    xmin=0, xmax=11.0,
    ymin=0, ymax=1.1
    ]
    
\pgfplotstableread{
 0.0625	1.0002
0.1875	1.00015
0.3125	1.0001
0.4375	1.
0.5625	0.9998750000000001
0.6875	0.9997075
0.8125	0.9994875000000001
0.9375	0.9991975
1.0625	0.9988349999999999
1.1875	0.9983675
1.3125	0.9978025
1.4375	0.9971075
1.5625	0.9962775
1.6875	0.9952925
1.8125	0.9941475
1.9375	0.992775
2.0625	0.9912025
2.1875	0.989425
2.3125	0.9874025
2.4375	0.9851049999999999
2.5625	0.9825550000000001
2.6875	0.9796699999999999
2.8125	0.976485
2.9375	0.9729774999999999
3.0625	0.9691000000000001
3.1875	0.9648699999999999
3.3125	0.9602675
3.4375	0.955255
3.5625	0.949865
3.6875	0.9440575
3.8125	0.9377875
3.9375	0.9311
4.0625	0.923945
4.1875	0.9163725
4.3125	0.9083325
4.4375	0.8998325
4.5625	0.8908525
4.6875	0.8813575
4.8125	0.8714275
4.9375	0.8610175
5.0625	0.8501225
5.1875	0.8387925
5.3125	0.827005
5.4375	0.8147325000000001
5.5625	0.8021
5.6875	0.7889925
5.8125	0.7754925
5.9375	0.761585
6.0625	0.744875
6.1875	0.730155
6.3125	0.715105
6.4375	0.6997325
6.5625	0.6840925
6.6875	0.6680950000000001
6.8125	0.6518299999999999
6.9375	0.6352925
7.0625	0.6185700000000001
7.1875	0.6016625
7.3125	0.5845825
7.4375	0.5673474999999999
7.5625	0.5500225000000001
7.6875	0.532605
7.8125	0.51513
7.9375	0.49761500000000003
8.0625	0.480105
8.1875	0.46265
8.3125	0.4452425
8.4375	0.42791250000000003
8.5625	0.41068000000000005
8.6875	0.39358250000000006
8.8125	0.37672249999999996
8.9375	0.36003999999999997
9.0625	0.3436225
9.1875	0.3274075
9.3125	0.31149000000000004
9.4375	0.2959
9.5625	0.2806225
9.6875	0.26571750000000005
9.8125	0.25122
9.9375	0.2370775
10.062	0.2233175
10.188	0.21000000000000002
10.312	0.197085
10.438	0.184635
10.562	0.17263
10.688	0.161095
10.812	0.150085
10.938	0.1395075
11.062	0.1294225
11.188	0.11978
11.312	0.1106325
11.438	0.1019725
11.562	0.09381075
11.688	0.08610275
11.812	0.07883925
11.938	0.072033
}\PythiaDYtdistICT

\pgfplotstableread{
 0.0625	1.0005
0.1875	1.0004
0.3125	1.0002
0.4375	1.
0.5625	0.9997325
0.6875	0.999385
0.8125	0.99895
0.9375	0.9984175
1.0625	0.997765
1.1875	0.996985
1.3125	0.99605
1.4375	0.9949574999999999
1.5625	0.9936750000000001
1.6875	0.9922025
1.8125	0.9905075
1.9375	0.9885875
2.0625	0.986425
2.1875	0.98399
2.3125	0.9812575
2.4375	0.9782275
2.5625	0.97488
2.6875	0.9711925
2.8125	0.9671675000000001
2.9375	0.9628
3.0625	0.9580325
3.1875	0.95286
3.3125	0.9473075
3.4375	0.941325
3.5625	0.9349375
3.6875	0.92811
3.8125	0.920835
3.9375	0.9131525
4.0625	0.9050149999999999
4.1875	0.89642
4.3125	0.8873249999999999
4.4375	0.8777925
4.5625	0.86783
4.6875	0.8574225
4.8125	0.8465675
4.9375	0.8352649999999999
5.0625	0.8235375
5.1875	0.8113925
5.3125	0.79877
5.4375	0.7857624999999999
5.5625	0.7724025
5.6875	0.7585875
5.8125	0.74442
5.9375	0.72989
6.0625	0.7129125
6.1875	0.69773
6.3125	0.6821925
6.4375	0.6664475000000001
6.5625	0.6504175
6.6875	0.634145
6.8125	0.617705
6.9375	0.60107
7.0625	0.5842575
7.1875	0.5672975
7.3125	0.5502549999999999
7.4375	0.533135
7.5625	0.5160325
7.6875	0.49880749999999996
7.8125	0.4815425
7.9375	0.46440000000000003
8.0625	0.4472475
8.1875	0.430215
8.3125	0.413275
8.4375	0.396455
8.5625	0.3798325
8.6875	0.363425
8.8125	0.347225
8.9375	0.33128500000000005
9.0625	0.31557
9.1875	0.300145
9.3125	0.2850375
9.4375	0.27030750000000003
9.5625	0.25593
9.6875	0.2418975
9.8125	0.22825
9.9375	0.2149775
10.062	0.2021625
10.188	0.18980249999999999
10.312	0.177835
10.438	0.1663
10.562	0.155255
10.688	0.144605
10.812	0.1344875
10.938	0.1247875
11.062	0.11557
11.188	0.1068275
11.312	0.098525
11.438	0.0906855
11.562	0.0833045
11.688	0.076342
11.812	0.0698265
11.938	0.06370975000000001
}\PythiaDYtdistIHard

\pgfplotstableread{
 0.0625	1.0000499999999999
0.1875	1.
0.3125	1.
0.4375	1.
0.5625	0.99996
0.6875	0.9999025
0.8125	0.99982
0.9375	0.9997050000000001
1.0625	0.9995575000000001
1.1875	0.9993625
1.3125	0.9991025
1.4375	0.9987725000000001
1.5625	0.9983675000000001
1.6875	0.9978549999999999
1.8125	0.9972425
1.9375	0.996505
2.0625	0.9956225000000001
2.1875	0.9945775
2.3125	0.99336
2.4375	0.9919450000000001
2.5625	0.9903325000000001
2.6875	0.98849
2.8125	0.9863975
2.9375	0.98403
3.0625	0.981395
3.1875	0.9784550000000001
3.3125	0.9752025
3.4375	0.9716275
3.5625	0.9676974999999999
3.6875	0.963395
3.8125	0.9587275
3.9375	0.95366
4.0625	0.94818
4.1875	0.942245
4.3125	0.935905
4.4375	0.9291275000000001
4.5625	0.9218925
4.6875	0.9142175
4.8125	0.9060999999999999
4.9375	0.8974825
5.0625	0.888375
5.1875	0.8788024999999999
5.3125	0.8687875
5.4375	0.858285
5.5625	0.847315
5.6875	0.8358775
5.8125	0.8239575
5.9375	0.8116350000000001
6.0625	0.79471
6.1875	0.7814775
6.3125	0.76788
6.4375	0.753865
6.5625	0.739455
6.6875	0.7246524999999999
6.8125	0.7095275
6.9375	0.6940775
7.0625	0.678265
7.1875	0.6622175
7.3125	0.645885
7.4375	0.6293424999999999
7.5625	0.61252
7.6875	0.595515
7.8125	0.5783324999999999
7.9375	0.56107
8.0625	0.543605
8.1875	0.5260575000000001
8.3125	0.5084975
8.4375	0.4909025
8.5625	0.473345
8.6875	0.45581499999999997
8.8125	0.43834249999999997
8.9375	0.42100250000000006
9.0625	0.4037625
9.1875	0.3866775
9.3125	0.369745
9.4375	0.353055
9.5625	0.3366225
9.6875	0.3204325
9.8125	0.3045175
9.9375	0.28894000000000003
10.062	0.273695
10.188	0.258795
10.312	0.2442825
10.438	0.2302125
10.562	0.2165225
10.688	0.20329
10.812	0.1905175
10.938	0.1781825
11.062	0.166325
11.188	0.15498
11.312	0.14406750000000001
11.438	0.133685
11.562	0.12376000000000001
11.688	0.114345
11.812	0.1054325
11.938	0.096972
}\PythiaDYtdistISoft

    \addplot[black,semithick] table {\PythiaDYtdistICT};
    \addlegendentry{CT14 PDF}
    \addplot[red,semithick] table {\PythiaDYtdistIHard};
    \addlegendentry{Hard PDF}
    \addplot[blue,semithick] table {\PythiaDYtdistISoft};
    \addlegendentry{Soft PDF}

  \end{axis}
\end{tikzpicture}

\fi

\end{center}
\caption{
No-splitting function, $\overline\Pi(t)$, versus shower time according to \textsc{Pythia}.
}
\label{fig:rhoI}
\end{figure}
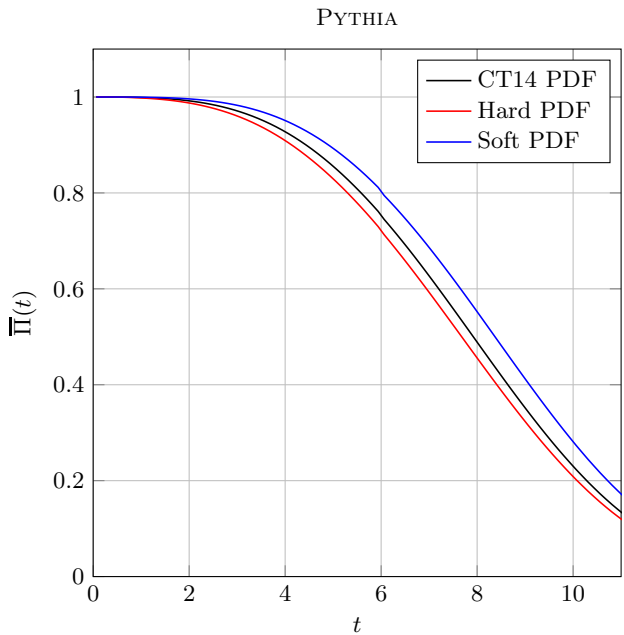

For a specified choice of parton variables, the probability for no splitting to occur before shower time $t$ is
\begin{equation}
\begin{split}
\Pi(x_\La x_\Lb s\, e^{-t}&, x_\La x_\Lb s; x_\La,a,x_\Lb,b)
\\&
= \int_t^\infty\!d\tau\ \rho(\tau;x_\La,a,x_\Lb,b)
\;.
\end{split}
\end{equation}
The average of this no-splitting probability is
\begin{equation}
\begin{split}
\label{eq:barPi}
\overline \Pi(t) ={}& \sum_{a,b}\int\!dx_\La \int\!dx_\Lb\
P(x_\La,a,x_\Lb,b)
\\&\times
\Pi(x_\La x_\Lb s\, e^{-t}, x_\La x_\Lb s; x_\La,a,x_\Lb,b)
\;,
\end{split}
\end{equation}
which equals
\begin{equation}
\overline \Pi(t)
= \int_t^\infty\!d\tau\ \bar\rho(\tau)
\;.
\end{equation}
We plot $\overline \Pi(t)$ for our three choices of PDF sets in Fig.~\ref{fig:rhoI}. We expect that at any fixed $t$, $\overline \Pi(t)$ will be smallest for the hard PDF set and largest for the soft PDF set. That is what we see. The differences between different PDF sets is substantial.

\begin{figure}
\begin{center}

\ifusefigs 

\begin{tikzpicture}
  \begin{axis}[
    title=\textsc{Pythia},
    xlabel={$t$}, ylabel={$\overline \Pi_\mathrm{alt}(t)$},
    legend cell align=left,
    every axis legend/.append style = {
      anchor=north east
    },
    xmin=0, xmax=11.0,
    ymin=0, ymax=1.1
    ]
    
\pgfplotstableread{
 0.0625	0.9963725
0.1875	0.988695
0.3125	0.98103
0.4375	0.9733525000000001
0.5625	0.965665
0.6875	0.957965
0.8125	0.9502325
0.9375	0.9424699999999999
1.0625	0.934655
1.1875	0.926785
1.3125	0.9188475
1.4375	0.91082
1.5625	0.9026975
1.6875	0.8944799999999999
1.8125	0.8861475
1.9375	0.8776599999999999
2.0625	0.869035
2.1875	0.8602825000000001
2.3125	0.8513675
2.4375	0.842265
2.5625	0.8330025
2.6875	0.82352
2.8125	0.8138575
2.9375	0.80399
3.0625	0.7938974999999999
3.1875	0.78359
3.3125	0.7730675
3.4375	0.7623025
3.5625	0.7513325
3.6875	0.74013
3.8125	0.728675
3.9375	0.717005
4.0625	0.705095
4.1875	0.6929875
4.3125	0.680655
4.4375	0.6681125
4.5625	0.6553575
4.6875	0.6423725
4.8125	0.6292175
4.9375	0.615875
5.0625	0.60235
5.1875	0.5886875
5.3125	0.5748825
5.4375	0.5609175
5.5625	0.5468875
5.6875	0.5327275
5.8125	0.5184925
5.9375	0.504185
6.0625	0.4882425
6.1875	0.47382749999999996
6.3125	0.45941
6.4375	0.44499750000000005
6.5625	0.430635
6.6875	0.41627000000000003
6.8125	0.4019575
6.9375	0.3877075
7.0625	0.3735675
7.1875	0.3595475
7.3125	0.34565
7.4375	0.3318975
7.5625	0.318315
7.6875	0.304915
7.8125	0.29171
7.9375	0.278715
8.0625	0.2659475
8.1875	0.2534425
8.3125	0.24119000000000002
8.4375	0.2291975
8.5625	0.2174825
8.6875	0.2060525
8.8125	0.194965
8.9375	0.18417250000000002
9.0625	0.17373
9.1875	0.16359
9.3125	0.1537975
9.4375	0.1443625
9.5625	0.13527
9.6875	0.1265375
9.8125	0.11817749999999999
9.9375	0.11015749999999999
10.062	0.10248
10.188	0.09517
10.312	0.088195
10.438	0.08157724999999999
10.562	0.07530224999999999
10.688	0.0693685
10.812	0.06379175
10.938	0.05852275
11.062	0.05358
11.188	0.0489345
11.312	0.04459475
11.438	0.040552000000000005
11.562	0.0367995
11.688	0.0333185
11.812	0.030089249999999998
11.938	0.02711125
}\PythiaDYtdistIaltCT

\pgfplotstableread{
 0.0625	0.9979325
0.1875	0.9927349999999999
0.3125	0.98751
0.4375	0.98224
0.5625	0.97691
0.6875	0.9715199999999999
0.8125	0.9660500000000001
0.9375	0.9604975
1.0625	0.9548449999999999
1.1875	0.94908
1.3125	0.9431849999999999
1.4375	0.9371575
1.5625	0.9309700000000001
1.6875	0.924615
1.8125	0.918085
1.9375	0.9113675
2.0625	0.904455
2.1875	0.8973175
2.3125	0.88995
2.4375	0.88234
2.5625	0.8744824999999999
2.6875	0.86636
2.8125	0.857985
2.9375	0.84935
3.0625	0.8404175
3.1875	0.83118
3.3125	0.8216725
3.4375	0.8118525000000001
3.5625	0.8017575
3.6875	0.7913525
3.8125	0.78064
3.9375	0.7696624999999999
4.0625	0.7583850000000001
4.1875	0.7468124999999999
4.3125	0.7349175
4.4375	0.7227525
4.5625	0.7103375000000001
4.6875	0.697665
4.8125	0.6847375
4.9375	0.6715574999999999
5.0625	0.6581575
5.1875	0.64454
5.3125	0.6306725
5.4375	0.6166324999999999
5.5625	0.6024375
5.6875	0.58804
5.8125	0.5734975
5.9375	0.5588200000000001
6.0625	0.5424
6.1875	0.52753
6.3125	0.5125474999999999
6.4375	0.49756
6.5625	0.4825125
6.6875	0.467445
6.8125	0.452415
6.9375	0.4374
7.0625	0.42242250000000003
7.1875	0.4075025
7.3125	0.39269
7.4375	0.37798
7.5625	0.36345249999999996
7.6875	0.3490025
7.8125	0.3346875
7.9375	0.32062250000000003
8.0625	0.30672
8.1875	0.29305749999999997
8.3125	0.27962
8.4375	0.266425
8.5625	0.2535175
8.6875	0.24091
8.8125	0.22859
8.9375	0.2165975
9.0625	0.2048975
9.1875	0.1935325
9.3125	0.1825175
9.4375	0.17187
9.5625	0.16159
9.6875	0.15166000000000002
9.8125	0.14209499999999997
9.9375	0.13288250000000001
10.062	0.12408000000000001
10.188	0.1156675
10.312	0.107605
10.438	0.09991225000000001
10.562	0.09261425000000001
10.688	0.08564574999999999
10.812	0.07908475
10.938	0.0728535
11.062	0.066989
11.188	0.061482749999999996
11.312	0.056297
11.438	0.0514385
11.562	0.04689925
11.688	0.04264325
11.812	0.038693
11.938	0.03498975
}\PythiaDYtdistIaltHard

\pgfplotstableread{
0.0625	0.995235
0.1875	0.98561
0.3125	0.9760125000000001
0.4375	0.96645
0.5625	0.956905
0.6875	0.94738
0.8125	0.937875
0.9375	0.9283825
1.0625	0.9188975
1.1875	0.9094175
1.3125	0.89992
1.4375	0.8904075
1.5625	0.8808625000000001
1.6875	0.8712774999999999
1.8125	0.86165
1.9375	0.85197
2.0625	0.84222
2.1875	0.8323900000000001
2.3125	0.8224725
2.4375	0.8124549999999999
2.5625	0.802335
2.6875	0.7921050000000001
2.8125	0.7817525000000001
2.9375	0.7712574999999999
3.0625	0.76064
3.1875	0.749865
3.3125	0.7389475000000001
3.4375	0.72787
3.5625	0.7166349999999999
3.6875	0.7052375000000001
3.8125	0.6936825
3.9375	0.6819675000000001
4.0625	0.67008
4.1875	0.65801
4.3125	0.6457925
4.4375	0.6334175
4.5625	0.6208899999999999
4.6875	0.6082125
4.8125	0.595415
4.9375	0.58246
5.0625	0.5693725000000001
5.1875	0.5561575000000001
5.3125	0.5428575
5.4375	0.52947
5.5625	0.5159925000000001
5.6875	0.502435
5.8125	0.4888075
5.9375	0.47516749999999996
6.0625	0.45908499999999997
6.1875	0.4454175
6.3125	0.43176499999999995
6.4375	0.41812249999999995
6.5625	0.4045125
6.6875	0.39094500000000004
6.8125	0.3774525
6.9375	0.36404000000000003
7.0625	0.35070500000000004
7.1875	0.3374975
7.3125	0.324415
7.4375	0.3114975
7.5625	0.298715
7.6875	0.2860975
7.8125	0.27368000000000003
7.9375	0.2615025
8.0625	0.2495
8.1875	0.23773249999999999
8.3125	0.2262225
8.4375	0.21497
8.5625	0.2039825
8.6875	0.1932875
8.8125	0.18286750000000002
8.9375	0.17274250000000002
9.0625	0.162925
9.1875	0.15342250000000002
9.3125	0.14424
9.4375	0.13538250000000002
9.5625	0.1268525
9.6875	0.1186425
9.8125	0.11076
9.9375	0.10322500000000001
10.062	0.09601625
10.188	0.0891295
10.312	0.082582
10.438	0.076374
10.562	0.0704735
10.688	0.06490125
10.812	0.059647500000000006
10.938	0.05469475
11.062	0.050033499999999995
11.188	0.045678500000000004
11.312	0.04160025
11.438	0.037799250000000006
11.562	0.034256
11.688	0.030973749999999998
11.812	0.02793975
11.938	0.025132500000000002
}\PythiaDYtdistIaltSoft

    \addplot[black,semithick] table {\PythiaDYtdistIaltCT};
    \addlegendentry{CT14 PDF}
    \addplot[red,semithick] table {\PythiaDYtdistIaltHard};
    \addlegendentry{Hard PDF}
    \addplot[blue,semithick] table {\PythiaDYtdistIaltSoft};
    \addlegendentry{Soft PDF}

  \end{axis}
\end{tikzpicture}

\fi

\end{center}
\caption{
Alternative no-splitting function, $\overline\Pi_\mathrm{alt}(t)$, versus shower time according to \textsc{Pythia}.
}
\label{fig:rhoIalt}
\end{figure}
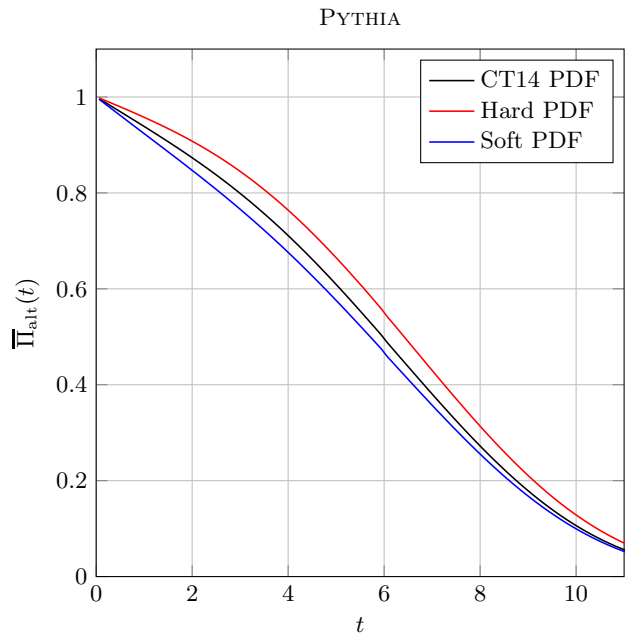

Now we can test the PDF property of initial state splittings. We define $\Pi_\textrm{alt}(\mu_\Ls^2, \mu_\scH^2; x_\La,a,x_\Lb,b)$ according to Eq.~(\ref{eq:Pialtdef}). Then in Fig.~\ref{fig:rhoIalt} we plot the average of $\Pi_\textrm{alt}(\mu_\Ls^2, \mu_\scH^2; x_\La,a,x_\Lb,b)$,
\begin{equation}
\begin{split}
\label{eq:barPiAlt}
\overline \Pi_\textrm{alt}(t) ={}& \sum_{a,b}\int\!dx_\La \int\!dx_\Lb\
P(x_\La,a,x_\Lb,b)
\\&\times
\Pi_\textrm{alt}(x_\La x_\Lb s\, e^{-t}, x_\La x_\Lb s; x_\La,a,x_\Lb,b)
\end{split}
\end{equation}
for our three choices for PDF sets. If the PDF property holds for a \textsc{Pythia} shower, the plots $\overline \Pi_\textrm{alt}(t)$ will be identical for the CT14 set, the soft set, and the hard set. For this to happen, $\overline \Pi_\textrm{alt}(t;\textrm{soft}) - \overline \Pi_\textrm{alt}(t;\textrm{CT14})$ needs to decrease compared to $\overline \Pi(t;\textrm{soft}) - \overline \Pi(t;\textrm{CT14})$ and $\overline \Pi_\textrm{alt}(t;\textrm{hard}) - \overline \Pi_\textrm{alt}(t;\textrm{CT14})$ needs to increase. This is what happens. However, the change is too pronounced, so that now 
\begin{equation}
\overline \Pi_\textrm{alt}(t;\textrm{soft})
< \overline \Pi_\textrm{alt}(t;\textrm{CT14})
< \overline \Pi_\textrm{alt}(t;\textrm{hard})
\;.
\end{equation}
We conclude that the PDF property does {\em not} hold for a \textsc{Pythia} shower in that the differences among the three PDF curves in Fig.~\ref{fig:rhoIalt} are not much smaller than the differences in Fig.~\ref{fig:rhoI}.

Should we have expected that the PDF property might not hold for a \textsc{Pythia} or for other parton shower Monte Carlo event generators? To see, we should look more closely at the derivation in Sec.~\ref{sec:PDFproperty}. We note, first of all, that the splitting function that governs an initial state splitting must be equal to $\widehat P(z)$ for a splitting that is close to the collinear limit, but that the parton splitting function will not generally equal $\widehat P(z)$ away from the collinear limit. Second, we note that the limits on the $z$ integrals in Eqs.~(\ref{eq:simplePi}), (\ref{eq:simplePipert}), and (\ref{eq:simplePiderivation}) are $z_-(\mu^2) < z < z_+(\mu^2)$. The derivation in Eq.~(\ref{eq:simplePiderivation}) treated $z_-(\mu^2)$ as being very close to 0 and $z_+(\mu^2)$ as being very close to 1. The upper limit, $z_+(\mu^2)$ is especially important. In an initial state splitting, the emitted gluon carries momentum fraction $1-z$. When $(1-z) \to 0$, the emitted gluon is becoming very soft. As explained in Ref.~\cite{pinkbook}, emissions with $(1-z) < (1-z_+(\mu^2))$ are omitted from the integration because they are ``unresolvable.'' But ``unresolvable'' means ``unresolvable at scale $\mu^2$.'' When $\mu^2$ is large, then $(1-z_+(\mu^2))$ is {\em not} small. 

Furthermore the integration region with small $(1-z)$ is important because the splitting function is singular for $(1-z) \to 0$ and the PDFs are fast varying when $(1-z)$ is small.

Just what the function $z_+(\mu^2)$ is depends on the choice of the hardness parameter $\mu^2$ used for ordering emissions in the shower and it depends on the definition of the kinematics used in the shower algorithm. But whatever the precise definitions are, we ought not to expect that $(1-z_+(\mu^2))$ is generally small. Thus we ought not to expect the PDF property to hold for initial state splittings in the shower.

\section{Test of the PDF property using Deductor}
\label{sec:deductortest}

We now examine the PDF property using our own parton shower event generator, \textsc{Deductor} \cite{Deductor}. This uses a dipole shower algorithm that is in many ways similar to that of \textsc{Pythia} \cite{pythia}. The default hardness variable $\mu^2$ in \textsc{Deductor} is a variable $\Lambda^2$ that is proportional to the virtuality of the splitting. However, \textsc{Deductor} has the option to use $k_\LT$ ordering. In this section, in order to stay reasonably close to the algorithm of \textsc{Pythia}, we use $k_\LT$ ordering. We note, however, the definition of $k_\LT^2$ in \textsc{Deductor} is not the same as the definition in \textsc{Pythia}. Some of the other differences between \textsc{Deductor} and \textsc{Pythia} are discussed in Ref.~\cite{NSThresholdII}.

Using \textsc{Deductor}, we start the initial state probability preserving parton shower evolution using a state with two incoming partons corresponding to a proton-proton collision with $\sqrt s = 13 \TeV$. The partons have momentum fractions $x_\La = x_\Lb = 0.2 = 2.6 \TeV/\sqrt{s}$. We choose parton flavors $a = \Lu$, $b = \bar \Lu$. In the notation of Sec.~\ref{sec:pythiatest}, the function $P(x_\La,a,x_\Lb,b)$ that gives the distribution of parton momentum fractions and flavors is simply a product of delta functions representing these choices.

\begin{figure}
\begin{center}

\ifusefigs 

\begin{tikzpicture}
  \begin{axis}[
    title= {\textsc{Deductor}, $k_\LT$ ordered},
    xlabel={$t$}, ylabel={$\overline \Pi_\mathrm{alt}(t)$},
    legend cell align=left,
    every axis legend/.append style = {
      anchor=north east
    },
    xmin=0, xmax=11.0,
    ymin=0, ymax=1.1
    ]
    
    \ratioplot[red,semithick]{fill=red!30!white, opacity=0.5}
    {HardDatakT.dat}{1}{3}{4}{7}{8}
    \addlegendentry{Hard PDF}
    
    \ratioplot[black,semithick]{fill=black!30!white, opacity=0.5}
    {CTDatakT.dat}{1}{3}{4}{7}{8}
    \addlegendentry{CT14 PDF}
    
    \ratioplot[blue,semithick]{fill=blue!30!white, opacity=0.5}
    {SoftDatakT.dat}{1}{3}{4}{7}{8}
    \addlegendentry{Soft PDF}

  \end{axis}
\end{tikzpicture}

\fi

\end{center}
\caption{
Alternative no-splitting function, $\overline\Pi_\mathrm{alt}(t)$, versus shower time according to \textsc{Deductor} with $k_\LT$ ordering.
}
\label{fig:ductrhoIalt}
\end{figure}
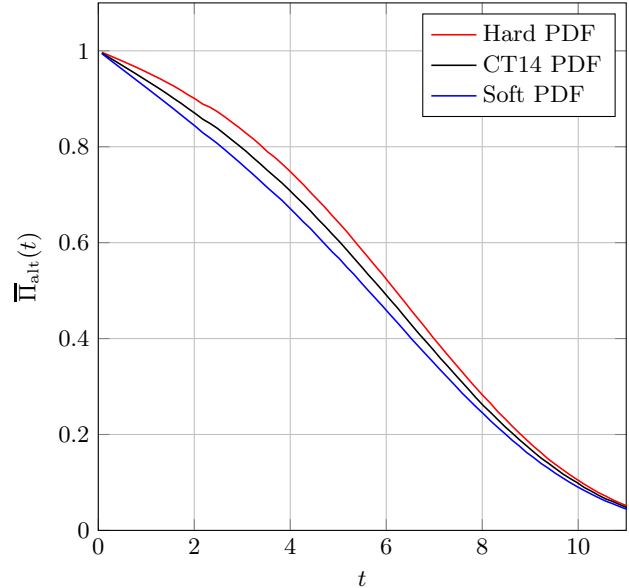

In Fig.~\ref{fig:ductrhoIalt}, we use the \textsc{Deductor} no-splitting operator and plot the alternative no-splitting function $\overline\Pi_\mathrm{alt}(t)$ as defined in Eqs.~(\ref{eq:Pialtdef}) and (\ref{eq:barPiAlt}) for our three choices of PDF sets. If the PDF property holds for the \textsc{Deductor} shower, the three curves should be identical. They are not. In fact, the three \textsc{Deductor} curves in Fig.~\ref{fig:ductrhoIalt} are quite similar to the three \textsc{Pythia} curves in Fig.~\ref{fig:rhoIalt}. 

We should not be surprised. The argument that the PDF property should not hold does not depend on features of a specific parton shower event generator. Rather, this reasoning applies quite generally.

\section{Restoring the missing physics}
\label{sec:threshold}

The results presented in the previous sections suggest that some physics may be missing in the shower evolution in \textsc{Pythia} and \textsc{Deductor}. In this section, we explore the missing physics. We describe an operator, in addition to the shower evolution operator, that is present in \textsc{Deductor}. We then provide numerical evidence that this operator provides the missing physics.

We need a certain amount of preparation to introduce the additional operator. For this, we follow the general theory of parton showers from Ref.~\cite{NSAllOrder}. The general theory works, in principle, at arbitrary order of perturbation theory. It recognizes that the splitting functions in a parton shower must match the infrared singularities of QCD but are rather arbitrary away from the singular regions. Nevertheless, the cross section corresponding to an infrared safe measurement should match between a perturbative QCD calculation and a calculation that uses a parton shower, up to the perturbative order available in both calculations. This matching requires a certain structure for the operators used in the parton shower calculation. Even when we use a shower with just order $\as$ splitting functions for the shower, the analysis suggests something about the structure needed in the parton shower calculation if we want to include potential large logarithms.

The theory we use for \textsc{Deductor} \cite{NSI, NSII, NSspin, NScolor, Deductor, ShowerTime, PartonDistFctns, NSbettercolor, NSexpipi, NSThreshold, NSAllOrder, NSThresholdII} uses a vector space, the ``statistical space,'' that describes the momenta, flavors, colors, and spins for all of the partons created in a shower as the shower develops.\footnote{The general theory includes parton spins but \textsc{Deductor} simply averages over spins.} The shower is then described using linear operators $\cU_\mathrm{pert}(\mu_\Ls^2,\mu_\Lh^2)$, $\cU(\mu_\Ls^2,\mu_\Lh^2)$, and $\cU_\cV(\mu_\Ls^2,\mu_\Lh^2)$ that act on this space. We describe these operators very briefly in this section, before turning to numerical results that illustrate their practical effects on the no-splitting probability. For readers who would like more details, we review the most relevant formulas from our previous papers in Appendix \ref{sec:operators}. We recommend reading the main text of this paper first, then referring to Appendix \ref{sec:operators}.

Following Ref.~\cite{NSAllOrder}, we begin with an operator $\cU_\mathrm{pert}(\mu_\Ls^2,\mu_\Lh^2)$ that describes evolution from a harder scale $\mu_\Lh^2$ to a softer scale $\mu_\Ls^2$. The operator $\cU_\mathrm{pert}(\mu_\Ls^2,\mu_\Lh^2)$ is constructed from Feynman diagrams with approximations that capture their collinear and soft limits. PDFs do not appear in $\cU_\mathrm{pert}(\mu_\Ls^2,\mu_\Lh^2)$. 

 The full shower evolution operator is constructed \cite{NSAllOrder} from $\cU_\mathrm{pert}(\mu_\Ls^2,\mu_\Lh^2)$ according to
\begin{align}
\label{eq:Udef}
&\cU(\mu_\Ls^2,\mu_\Lh^2) 
\\&= 
\cU_\cV(\mu_\Lf^2,\mu_\Ls^2)\,\cF(\mu_\Ls^2)\,
\cU_\mathrm{pert}(\mu_\Ls^2,\mu_\Lh^2)
\cF^{-1}(\mu_\Lh^2)\,\cU_\cV^{-1}(\mu_\Lf^2,\mu_\Lh^2)
\,.
\notag
\end{align}
There are two kinds of operators here that modify $\cU_\mathrm{pert}(\mu_\Ls^2,\mu_\Lh^2)$. 

The first is $\cF(\mu^2)$. This operator multiplies by PDFs $f_{a/A}(x_\La, \mu^2) f_{b/B}(x_\Lb, \mu^2)$ and divides by ${n_\Lc(a) n_\Ls(a)\, n_\Lc(b) n_\Ls(b)\ 4 x_\La x_\Lb s}$, where $n_\Lc(i)$ is the number of colors associated with a parton of flavor $i$ (3 or 8) and $n_\Ls(i)$ is the corresponding number of spin states (2). We apply the operator $\cF(\mu_\Ls^2)$ to the parton state after $\cU_\mathrm{pert}(\mu_\Ls^2,\mu_\Lh^2)$ acts. We apply the inverse operator $\cF^{-1}(\mu_\Lh^2)$ to the parton state before $\cU_\mathrm{pert}(\mu_\Ls^2,\mu_\Lh^2)$ acts.

The second kind of operator is $\cU_\cV(\mu_\Lf^2,\mu^2)$. This operator leaves the number of partons, their flavors, and their momenta unchanged. It is a necessary part of the general formalism. We write it as
\begin{equation}
\label{eq:UVexponential}
\cU_\cV(\mu_\Lf^2,\mu^2)
=\mathbb{T} \exp\!\left(
\int_{\mu_\Lf^2}^{\mu^{2}}\!\frac{d\bar\mu^2}{\bar\mu^2}\,\cS_\cV(\bar\mu^2)
\right)
\;.
\end{equation}
One could expand this operator perturbatively and keep only the first terms, but the exponent contains large logarithms, so we use the exponential form. When we use a first order shower, we evaluate $\cS_\cV$ up to order $\as$. The scale $\mu_\Lf^2$ is an infrared cutoff scale of order $1 \GeV^2$. The integrand in the exponent is convergent in the infrared if $\as$ is held constant, so the result is not sensitive to the choice of $\mu_\Lf^2$. An approximation to $\cS_\cV(\mu^2)$ at first order is part of \textsc{Deductor}. It contains perturbative factors and ratios of PDFs.

The operator $\cU_\cV(\mu_\Lf^2,\mu^2)$ makes the complete shower evolution operator $\cU(\mu_\Ls^2,\mu_\Lh^2)$ probability preserving. In the notation \cite{NSI} that we use, the total probability associated with a statistical state $\sket{\rho}$ is obtained by summing over the number of partons, integrating over momenta, summing over flavors, and taking the trace in color and in spin. This total probability is denoted as $\sbrax{1}\sket{\rho}$. With the $\cU_\cV$ factors included, one has
\begin{equation}
\sbra{1}\cU(\mu_\Ls^2,\mu_\Lh^2) \sket{\rho} = \sbrax{1}\sket{\rho}
\;,
\end{equation}
so that the shower does not change the total probability. The showers in \textsc{Pythia} and other parton shower event generators are constructed to be probability preserving. We will discuss the method for arranging this at the end of this section. 

The operator $\cU_\cV(\mu_\Lf^2,\mu^2)$ is also numerically important. It describes threshold logarithms \cite{Sterman1987, Catani32, CMW, CataniManganoNason32, SudakovFactorization, ManoharSCET, BecherNeubertPecjak, Stewart:2009yx} in the context of a parton shower.\footnote{See also the more extensive list of papers about threshold logarithms in Ref.~\cite{NSThresholdII}.} We have analyzed $\cU_\cV(\mu_\Lf^2,\mu^2)$ in Refs.~\cite{NSThreshold, NSThresholdII}.

We are now prepared to examine the probability not to have a splitting between two scales in the \textsc{Deductor} shower. The operator $\cU(\mu^2,\mu_\Lh^2)$ obeys an evolution equation of the form
\begin{equation}
\begin{split}
\label{eq:Uevolution}
\mu^2\frac{d}{d\mu^2}\,&\cU(\mu^2,\mu_\Lh^2)
\\&
= - [\cS_\mathrm{split}(\mu^2) + \cS_\textrm{no-split}(\mu^2)]\,
\cU(\mu^2,\mu_\Lh^2)
\;.%
\end{split}
\end{equation}
Here $\cS_\mathrm{split}(\mu^2)$ increases the number of partons and $\cS_\textrm{no-split}(\mu^2)$ leaves the number of partons unchanged. This evolution equation can be solved in the form
\begin{align}
\label{eq:Usoln}
\cU(\mu_\Ls^2,\mu_\Lh^2) ={}& \cN(\mu_\Ls^2,\mu_\Lh^2)
\\&
+ \int_{\mu_\Ls^2}^{\mu_\Lh^2}\!\frac{d\mu^2}{\mu^2}\,
\cU(\mu_\Ls^2,\mu^2)
\cS_\mathrm{split}(\mu^2)
\cN(\mu^2,\mu_\Lh^2)
,
\notag
\end{align}
where
\begin{equation}
\begin{split}
\label{eq:Nfull}
\cN (\mu_\Ls^2,\mu_\Lh^2)
={}& \mathbb{T} \exp\!\left(
\int_{\mu_\Ls^2}^{\mu_\Lh^2}\!\frac{d\mu^2}{\mu^2}\,
\cS_{\textrm{no-split}}(\mu^2)
\right)
.
\end{split}
\end{equation}
Here $\mathbb{T}$ indicates ordering of the exponential in the scale $\mu^2$, with lower scales to the left. The operator $\cN(\mu_\Ls^2,\mu_\Lh^2)$ is the no-splitting operator. The quantity $\sbra{1}\cN(\mu_\Ls^2,\mu_\Lh^2)\sket{\rho}$ gives the probability for the state $\sket{\rho}$ to evolve from scale $\mu_\Lh^2$ to scale $\mu_\Ls^2$ without splitting. Then Eq.~(\ref{eq:Usoln}) says that either the shower can evolve from scale $\mu_\Lh^2$ to scale $\mu_\Ls^2$ without splitting or it can evolve to an intermediate scale $\mu^2$ without splitting, then undergo a splitting at scale $\mu^2$, then evolve from $\mu^2$ to $\mu_\Ls^2$ with zero or more splittings. 

The operator $\cN (\mu_\Ls^2,\mu_\Lh^2)$ leaves the number of partons and their momenta and flavors unchanged. Thus we call it the no-splitting operator. However, it has a nontrivial action on the parton colors. In order to deal with the complications of quantum color, \textsc{Deductor} starts with an approximate treatment of color, the LC+ approximation \cite{NScolor}. Then color beyond this approximation can be included perturbatively \cite{NSbettercolor}. In this paper, we start with a quark-antiquark pair in a color singlet state and consider the probability to have no emissions between two scales. For this problem, the needed color treatment is trivial. However, a note of caution is in order: one should not imagine that the color content of $\cN (\mu_\Ls^2,\mu_\Lh^2)$ is always trivial.

The operator $\cU_\mathrm{pert}(\mu^2,\mu_\Lh^2)$ obeys a similar equation,
\begin{align}
\label{eq:Upertevolution}
\mu^2\frac{d}{d\mu^2}\,&\cU_\mathrm{pert}(\mu^2,\mu_\Lh^2)
\\&
= - [\cS_\mathrm{split}^\mathrm{pert}(\mu^2) 
+ \cS_\textrm{no-split}^\mathrm{pert}(\mu^2)]\,
\cU_\mathrm{pert}(\mu^2,\mu_\Lh^2)
\;.
\notag
\end{align}
The operators $\cS_\mathrm{split}^\mathrm{pert}(\mu^2)$ and $\cS_{\textrm{no-split}}^\mathrm{pert}(\mu^2)$ are calculated from the singular behavior of Feynman diagrams. They do not involve PDFs. At first order, $\cS_\mathrm{split}^\mathrm{pert}(\mu^2)$ comes from diagrams with a real emission, while $\cS_\textrm{no-split}^\mathrm{pert}(\mu^2)$ corresponds to virtual graphs. 

The first order version of $\cS_\textrm{no-split}^\mathrm{pert}(\mu^2)$ used in \textsc{Deductor} is approximate in that it omits the contributions from the imaginary part of virtual graphs. These contributions can be included perturbatively \cite{NSbettercolor} or even in exponentiated form \cite{NSexpipi}, but we find that they are not numerically important \cite{NSbettercolor, NSexpipi}.

This evolution equation can be solved in the same form as above,
\begin{align}
\label{eq:Upertsoln}
\cU_\mathrm{pert}&(\mu_\Ls^2,\mu_\Lh^2) 
\notag
\\={}& 
\cN_\mathrm{pert}(\mu_\Ls^2,\mu_\Lh^2)
\\&
+ \int_{\mu_\Ls^2}^{\mu_\Lh^2}\!\frac{d\mu^2}{\mu^2}\,
\cU_\mathrm{pert}(\mu_\Ls^2,\mu^2)
\cS_\mathrm{split}^\mathrm{pert}(\mu^2)
\cN_\mathrm{pert}(\mu^2,\mu_\Lh^2)
\;.
\notag
\end{align}
Here the perturbative no-splitting operator is
\begin{equation}
\begin{split}
\label{eq:Npert}
\cN_\mathrm{pert}(\mu_\Ls^2,\mu_\Lh^2)
={}&
\mathbb{T} \exp\!\left(
\int_{\mu_\Ls^2}^{\mu_\Lh^2}\!\frac{d\mu^2}{\mu^2}\,
\cS_{\textrm{no-split}}^\mathrm{pert}(\mu^2)
\right)
.
\end{split}
\end{equation}

Now we can use the relation (\ref{eq:Udef}) between $\cU(\mu_\Ls^2,\mu_\Lh^2)$  and $\cU_\mathrm{pert}(\mu_\Ls^2,\mu_\Lh^2)$. Using this relation in Eqs.~(\ref{eq:Usoln}) and (\ref{eq:Upertsoln}), we see that the no-splitting operators are related by
\begin{align}
\label{eq:Nrelations}
&\cN(\mu_\Ls^2,\mu_\Lh^2) 
\\&= 
\cU_\cV(\mu_\Lf^2,\mu_\Ls^2)\,\cF(\mu_\Ls^2)\,
\cN_\mathrm{pert}(\mu_\Ls^2,\mu_\Lh^2)
\cF^{-1}(\mu_\Lh^2)\,\cU_\cV^{-1}(\mu_\Lf^2,\mu_\Lh^2)
\,.
\notag
\end{align}
Then the corresponding generators of splittings are related by
\begin{align}
\label{eq:Ssplitrelations}
\cS_\mathrm{split}&(\mu^2) 
\\&= 
\cU_\cV(\mu_\Lf^2,\mu^2)\,\cF(\mu^2)\,
\cS_\mathrm{split}^\mathrm{pert}(\mu^2)
\cF^{-1}(\mu^2)\,\cU_\cV^{-1}(\mu_\Lf^2,\mu^2)
\,.
\notag
\end{align}
This relation is simpler at lowest perturbative order. The perturbative expansions of both $\cS_\mathrm{split}(\mu^2)$ and $\cS_\mathrm{split}^\mathrm{pert}(\mu^2)$ begin at order $\as$, while the perturbative expansion of $\cU_\cV(\mu_\Lf^2,\mu_\Ls^2)$ begins with $\cU_\cV(\mu_\Lf^2,\mu_\Ls^2) = 1 + \cO(\as)$. Thus the relation that connects the splitting functions used in a first order parton shower is
\begin{equation}
\label{eq:Ssplitrelations}
\cS_\mathrm{split}^{(1)}(\mu^2) 
= 
\cF(\mu^2)\,
\cS_\mathrm{split}^\mathrm{pert,(1)}(\mu^2)\,
\cF^{-1}(\mu^2)
\,.
\end{equation}
An initial state splitting comes with the ratio of the PDFs after the splitting to the PDFs before the splitting. This, of course, is the standard form for a parton shower with backward evolution for the initial state partons.

The relation (\ref{eq:Nrelations}) is of special interest for our study of the PDF property. We have, after noting that $\cF$ commutes with $\cU_\cV$, 
\begin{equation}
\begin{split}
\label{eq:Nrelationsreversed}
\cU_\cV^{-1}(\mu_\Lf^2,\mu_\Ls^2)&\,\cF^{-1}(\mu_\Ls^2)\,
\cN(\mu_\Ls^2,\mu_\Lh^2)\,
\cF(\mu_\Lh^2)\,\cU_\cV(\mu_\Lf^2,\mu_\Lh^2)
\\&=
\cN_\mathrm{pert}(\mu_\Ls^2,\mu_\Lh^2)
\,.
\end{split}
\end{equation}
Recall that the operator $\cN_\mathrm{pert}(\mu_\Ls^2,\mu_\Lh^2)$ is constructed from perturbative Feynman diagrams but does not involve PDFs. Therefore the operator on the left-hand side of Eq.~(\ref{eq:Nrelationsreversed}) is unchanged if we change PDFs.

Using the operator notation, the alternative no-splitting function is
\begin{equation}
\overline \Pi_\textrm{alt}(t) = 
\sbra{1}
\cF^{-1}(\mu_\Ls^2)\,
\cN(\mu_\Ls^2,\mu_\scH^2)
\cF(\mu_\scH^2)
\sket{\rho_\scH}
\;,
\end{equation}
where $\sket{\rho_\scH}$ is the initial partonic statistical state at scale $\mu_\scH^2$. This function can depend on the PDF set used in its calculation. As we saw in Fig.~\ref{fig:ductrhoIalt}, $\overline \Pi_\textrm{alt}(t)$ does indeed depend on the PDF set used. 

Because of Eq.~(\ref{eq:Nrelationsreversed}) and because $\cN_\mathrm{pert}(\mu_\Ls^2,\mu_\Lh^2)$ does not involve PDFs, the function
\begin{equation}
\begin{split}
\label{eq:Pialtmod}
\overline \Pi_{\textrm{opt}}(t) 
={}& 
\sbra{1}
\cU_\cV^{-1}(\mu_\Lf^2,\mu_\Ls^2)\cF^{-1}(\mu_\Ls^2)
\\&\times
\cN(\mu_\Ls^2,\mu_\scH^2)
\cF(\mu_\scH^2)\,\cU_\cV(\mu_\Lf^2,\mu_\scH^2)
\sket{\rho_\scH}
\,
\end{split}
\end{equation}
is independent of the PDF set used.\footnote{We also note that there is no reason that either $\overline \Pi_{\textrm{alt}}(t)$ or $\overline \Pi_{\textrm{opt}}(t)$ should be everywhere smaller than 1, since these quantities are not probabilities.} (Here ``opt'' is for ``optimal.'') \textsc{Deductor} contains an approximate version of the operator $\cU_\cV(\mu_\Lf^2,\mu^2)$. Using this operator, we can directly check the dependence of $\overline \Pi_{\textrm{opt}}(t)$ on the parton distribution set used. The result is shown in Fig.~\ref{fig:ductrhoIaltV}.

\begin{figure}
\begin{center}

\ifusefigs 

\begin{tikzpicture}
  \begin{axis}[
    title= {\textsc{Deductor}, $k_\LT$ ordered},
    xlabel={$t$}, ylabel={$\overline \Pi_{\mathrm{opt}}(t)$},
    legend cell align=left,
    every axis legend/.append style = {
      anchor=north east
    },
    xmin=0, xmax=11.0,
    ymin=0, ymax=1.1
    ]
    
    \ratioplot[red,semithick]{fill=red!30!white, opacity=0.5}
    {HardDatakT.dat}{1}{5}{6}{7}{8}
    \addlegendentry{Hard PDF}
    
    \ratioplot[black,semithick]{fill=black!30!white, opacity=0.5}
    {CTDatakT.dat}{1}{5}{6}{7}{8}
    \addlegendentry{CT14 PDF}
    
    \ratioplot[blue,semithick]{fill=blue!30!white, opacity=0.5}
    {SoftDatakT.dat}{1}{5}{6}{7}{8}
    \addlegendentry{Soft PDF}

  \end{axis}
\end{tikzpicture}

\fi

\end{center}
\caption{
Modified no-splitting fraction $\overline \Pi_{\mathrm{opt}}(t)$, including the threshold operator $\cU_\cV(\mu_\Lf^2,\mu^2)$, versus shower time according to \textsc{Deductor} with $k_\LT$ ordering.
}
\label{fig:ductrhoIaltV}
\end{figure}

In \textsc{Deductor}, there are some approximations in both $\cN(\mu_\Ls^2,\mu_\scH^2)$ and in $\cU_\cV(\mu_\Lf^2,\mu^2)$, so we cannot expect $\overline \Pi_{\textrm{opt}}(t)$ as calculated to be exactly independent of the PDFs. Indeed, the three curves for the CT14, hard, and soft PDF sets are not exactly the same. Nevertheless, the agreement is quite good. 

If we assume that the dependence on the parton distributions seen in $\overline \Pi_{\textrm{alt}}(t)$, Fig.~\ref{fig:ductrhoIalt}, is a sign that some physics is missing from its definition, then Fig.~\ref{fig:ductrhoIaltV} suggests that the missing physics is contained in $\cU_\cV(\mu_\Lf^2,\mu^2)$.

\section{Discussion}
\label{sec:discussion}

Within the framework of Ref.~\cite{NSAllOrder}, the shower evolution operator $\cU(\mu_\Ls^2,\mu_\Lh^2)$ is probability preserving thanks to the inclusion of the operator $\cU_\cV(\mu_\Lf^2,\mu^2)$ in its definition (\ref{eq:Udef}). In \textsc{Pythia} and other parton shower event generators, the parton shower is probability preserving without the need for $\cU_\cV(\mu_\Lf^2,\mu^2)$. The method for guaranteeing this property is simple. Consider the case of just one hadron and just one kind of parton as in Sec.~\ref{sec:PDFproperty}. When the initial state parton has momentum fraction $x$, the differential probability for the initial state parton  to split at scale $\mu^2$ with momentum fraction $z$ and azimuthal angle $\phi$ is
\begin{equation}
\label{eq:splittingprobability}
\rho(\mu^2, z, \phi) = 
\frac{\as(\mu^2)}{2\pi}\,
G(\mu^2, z, \phi)\,
\frac{f(x/z, \mu^2)}{f(x,\mu^2)}
\;,
\end{equation}
where $G(\mu^2, z, \phi)$ is the splitting function, approximately equal to the DGLAP splitting kernel $\widehat P(z)$. The splitting function vanishes for $z > z_+(\mu^2)$, where $z_+(\mu^2)$ is determined by the definition of the ordering variable and the shower kinematics. 

The shower algorithm needs a function giving the probability for the initial state parton not to split between a scale $\mu_\Lh^2$ and a lower scale $\mu_\Ls^2$. The standard method is to define this function to be
\begin{equation}
\begin{split}
\label{eq:simplePimod}
\Pi(\mu_\Ls^2, \mu_\Lh^2, x) ={}& 
\exp\!\bigg(
- \int_{\mu_\Ls^2}^{\mu_\Lh^2}\!\frac{d\mu^2}{\mu^2}
\int\!\frac{dz}{z} \int\!d\phi
\\&\times 
\frac{\as(\mu^2)}{2\pi}\, 
G(\mu^2, z, \phi)\,
\frac{f(x/z, \mu^2)}{f(x,\mu^2)}
\bigg)
,
\end{split}
\end{equation}
as in Eq.~(\ref{eq:simplePi}). The exponent is the negative of the inclusive probability to have a splitting, so the exponential is the probability not to have a splitting. This makes the shower probability preserving.

We should note, however, that if we base the parton shower algorithm purely on the properties of perturbative QCD, then for a lowest order shower we have available real emission diagrams and one loop virtual diagrams. We then start with a perturbative no-splitting operator $\cN_\mathrm{pert}(\mu_\Ls^2,\mu_\Lh^2)$ as given in Eq.~(\ref{eq:Npert}). In the exponent of $\cN_\mathrm{pert}(\mu_\Ls^2,\mu_\Lh^2)$, the operator $\cS_{\textrm{no-split}}^\mathrm{pert}(\mu^2)$ is derived from one loop virtual diagrams. Since we know that virtual diagrams cancel the completely inclusive integral of real emission diagrams, Eq.~(\ref{eq:simplePimod}) is a plausible approximation for the effect of the virtual diagrams. However, Eq.~(\ref{eq:simplePimod}) cannot be exact because the virtual diagrams do not contain the ratio of PDFs in Eq.~(\ref{eq:simplePimod}). In order to connect the evolution of the parton shower to the evolution of the PDFs, we need something like the approximations used in Sec.~\ref{sec:intro} to give us the PDF property. Thus, a good way to test the approximations that give us Eq.~(\ref{eq:simplePimod}) is to test the PDF property. That is what we have done in the previous sections. This test can be applied quite easily to any parton shower event generator.

What is the practical effect of $\cU_\cV(\mu_\Lf^2,\mu^2)$? The result of Ref.~\cite{NSAllOrder} is that the cross section corresponding to an infrared safe observable $J$ is given by
\begin{equation}
\begin{split}
\label{eq:sigmaU8}
\sigma[J] ={}& 
\sbra{1} \cO_J\, 
\cU(\mu_\Lf^2,\mu_\scH^2)\,
\cU_\cV(\mu_\Lf^2,\mu_\scH^2)\,
\cF(\mu_\scH^2)
\sket{\rho_\scH}
\\&
+\cO(\as^{k+1}) + \cO(\mu_\Lf^2/Q[J]^2)
\;.
\end{split}
\end{equation}
This cross section is determined by standard QCD perturbation theory and is independent (up to the perturbative order calculated) of the many choices used in the definition of the shower. There is a perturbative error term $\cO(\as^{k+1})$ coming from uncalculated higher order corrections.\footnote{The operators $\cU(\mu_\Lf^2,\mu_\scH^2)$ and $\cU_\cV(\mu_\Lf^2,\mu_\scH^2)$ include contributions at all orders of perturbation theory, but the strict perturbative accuracy of Eq.~(\ref{eq:sigmaU8}) is limited to the accuracy at which the hard scattering, $\sket{\rho_\scH}$, is calculated.} There is also a power suppressed error term $\cO(\mu_\Lf^2/Q[J]^2)$ that is small if the infrared cutoff scale $\mu_\Lf^2$ is much smaller than the scale $Q[J]^2$ corresponding to the observable that we measure. There is an operator  $\cO_J$ that specifies the observable to be measured, followed by a statistical bra state $\sbra{1}$ that instructs us to integrate over all parton variables. There is a statistical state $\sket{\rho_\scH}$ corresponding to a hard scattering at scale $\mu_\scH^2$. If we calculate beyond leading order, then $\sket{\rho_\scH}$ includes appropriate infrared subtractions. There is an operator $\cF(\mu_\scH^2)$ that inserts the PDFs for the hard scattering. 

There is the operator $\cU(\mu_\Lf^2,\mu_\scH^2)$ that generates a probability preserving shower from the hard scale $\mu_\scH^2$ to the shower cutoff scale $\mu_\Lf^2$. At lowest order, and ignoring for simplicity the imaginary parts of virtual diagrams, the no-splitting operator in the shower is related to the shower splitting functions by the analogue of the simple relation (\ref{eq:simplePimod}). So far, the lowest order version of this is very standard. 

Finally there is the operator $\cU_\cV(\mu_\Lf^2,\mu_\scH^2)$. As we have noted, this operator is needed to make the shower operator $\cU(\mu_\Lf^2,\mu_\scH^2)$ probability preserving. It also allows the cross section $\sigma[J]$ to be independent, up to the perturbative order calculated, of the choices made in defining the shower and the subtractions in $\sket{\rho_\scH}$. The operator $\cU_\cV(\mu_\Lf^2,\mu_\scH^2)$ appears immediately after the statistical state representing the hard scattering. 

Examination of $\cU_\cV(\mu_\Lf^2,\mu_\scH^2)$ \cite{NSThreshold, NSThresholdII} shows that, with $k_\LT$ ordering, this operator generates a summation of threshold logarithms corresponding to the hard scattering.

Threshold logarithms have been understood from an analytical perspective for a long time \cite{Sterman1987, Catani32, CMW, CataniManganoNason32, SudakovFactorization, ManoharSCET, BecherNeubertPecjak, Stewart:2009yx} and are known to have a substantial effect on cross sections in many important cases. Thus it is not a surprise that we see substantial differences between Fig.~\ref{fig:ductrhoIalt} and Fig.~\ref{fig:ductrhoIaltV}.

\section{Deductor with $\Lambda$ or angle ordering}
\label{sec:otherordering}

In Secs. \ref{sec:deductortest} and \ref{sec:threshold}, \textsc{Deductor} results were obtained with $k_\LT$ ordering because that is most similar to the ordering used in \textsc{Pythia}. However, the default ordering variable in \textsc{Deductor} is $\Lambda^2$, which is proportional to the virtuality in a splitting. \textsc{Deductor} also has the option to use angle ordering. In this section, we explore the PDF property using these other ordering variables. We will find that the ordering matters.

\subsection{Scale variables}
\label{sec:scales}

Consider a splitting of an initial state parton with momentum $p_\La$ into a new initial state parton with momentum $\hat p_\La$ and a new final state parton with momentum $\hat p_{m+1}$. Let $Q_0$ denote a fixed vector equal to the total momentum of the final state partons just after the hard scattering that initiates the shower and let $Q$ denote the total momentum of the final state partons just before the splitting. We take all partons to be massless. Let $\mu^2_1$ denote the virtuality in the splitting,
\begin{equation}
\mu^2_1 = - (\hat p_\La - \hat p_{m+1})^2
\;.
\end{equation}
We also define the momentum fraction in the splitting to be
\begin{equation}
z = \frac{p_\La \cdot p_\Lb}{\hat p_\La \cdot p_\Lb}
\;.
\end{equation}
Here $p_\Lb$ is the momentum of the other initial state parton. With this notation, the default ordering variable $\Lambda^2$ is defined for an initial state splitting by
\begin{equation}
\label{eq:LambdaDef}
\Lambda^2 = \mu^2_1\, \frac{Q_0^2}{2 p_\La \cdot Q_0}
\;.
\end{equation}

For this same splitting, we define a transverse momentum variable by
\begin{equation}
\bm k_\LT^2 = (1-z)\, \mu^2_1
\end{equation}
and use this as the ordering variable for a $k_\LT$-ordered shower. It is notationally convenient to define a corresponding scale variable by $\mu_0^2 = \bm k_\perp^2$. Then
\begin{equation}
\mu_0^2 = (1-z)\, \mu^2_1
\;.
\end{equation}

We also define an angle variable by
\begin{equation}
\bm \theta^2 = \frac{\mu^2_1}{(1-z)}\,\frac{Q_0^2}{(p_\La\cdot Q_0)^2}
\;.
\end{equation}
\textsc{Deductor} takes the ordering variable for an angle ordered shower to be 
$\mu^2_2 = \bm \theta^2 (p_\La\cdot Q_0)^2/Q_0^2$. Then
\begin{equation}
\mu^2_2 = \frac{\mu^2_1}{1-z}
\;.
\end{equation}
We could also use $\bm \theta^2 = \mu_2^2\,Q_0^2/(p_\La\cdot Q_0)^2$ directly as the ordering variable. This change would not affect the numerical results later in this section.

We now have scale variables $\mu_0^2$, $\mu_1^2$ and $\mu_2^2$ for $k_T$-ordered, $\Lambda$-ordered, and angle-ordered showers respectively. These are related by
\begin{equation}
\label{eq:musubstitution}
\mu_0^2 = (1-z)^\lambda \mu_\lambda^2
\;,
\hskip 1 cm 
\lambda = 0,1,2
\;.
\end{equation}

\subsection{Parton distribution functions}
\label{sec:PDFs}

Now consider the evolution of the PDFs, following the argument that we have used for $\Lambda$ ordering \cite{NSThreshold}. We write the first order DGLAP kernels that apply to $\MSbar$ PDFs in the form
\begin{equation}
\label{Pregdef}
\widehat P_{a\hat a}(z) = \delta_{a \hat a}\,\frac{2 z C_a}{1-z}
+ P_{a\hat a}^{\rm reg}(z)
\;,
\end{equation}
where $P_{a\hat a}^{\rm reg}(z)$ is nonsingular as $z \to 1$. Here $C_\Lg = C_\LA$ and $C_q = C_\LF$ for $q\in \{\Lu,  \bar \Lu, \Ld, \dots\}$. We also define standard constants $\gamma_a$ by
\begin{equation}
\begin{split}
\label{eq:gammaf}
\gamma_\Lg ={}& \frac{11 C_\LA}{6} - \frac{2 T_\LR N_\Lf}{3}
\;,
\\
\gamma_q ={}& \frac{3 C_\LF}{2}\;, \qquad q\in \{\Lu,  \bar \Lu, \Ld, \dots\}
\;.
\end{split}
\end{equation}
Finally we include a standard factor \cite{CMW}
\begin{equation}
\label{eq:CMW}
\lambda_\LR = \exp\left(- \frac{  C_\LA(67 - 3\pi^2)- 10\, n_\Lf}
{3\, (33 - 2\,n_\Lf)}\right)
\end{equation}
in the argument of $\as$ in the evolution equation for the PDFs.

Then we can write the DGLAP evolution equation in a form that implements the + prescription as
\begin{equation}
\begin{split}
\label{eq:DGLAP0alt}
\mu_0^2 &\frac{d f_{a/A}^{\MSbar}(x,\mu_0^2)}{d\mu_0^2}
\\ 
={}& 
\frac{\as(\lambda_\LR \mu_0^2)}{2\pi}\,
\gamma_a\,
f_{a/A}^{\MSbar}(x,\mu_0^2)\,
\theta(\mu_0^2 > m_0^2(a))
\\&   
+\int_0^1\!dz\ \frac{\as(\lambda_\LR \mu_0^2)}{2\pi}\,
\theta(\mu_0^2 > m_0^2(a))
\\& \quad\times
\Bigg( 
\frac{2 C_a}{1-z}
\left[
f_{a/A}^{\MSbar}(x/z,\mu_0^2)
- f_{a/A}^{\MSbar}(x,\mu_0^2)
\right]
\\&\qquad
+\sum_{\hat a} \frac{1}{z}\,P_{a\hat a}^{\rm reg}(z)\,
f_{\hat a/A}^{\MSbar}(x/z,\mu_0^2) 
\Bigg)
\;.
\end{split}
\end{equation}
We have included a cut $\mu_0^2 > m_0^2(a)$. For light flavors $a$, we take $m_0^2(a)$ to be the scale, of order 1 GeV, at which we specify initial conditions for the PDFs. Then the cut turns off evolution for $\mu_0^2$ smaller than $m_0^2(a)$, so that we can impose the initial conditions at any smaller scale. For charm and bottom quarks, we define $m_0(a)$ to be the corresponding quark mass, so that evolution starts at this scale.

Now, the $\MSbar$ definition of the PDFs is that these functions are renormalized by subtracting poles in $\epsilon$, where loop integrals are performed with dimensional regularization: $\mu_0^{2\epsilon} \int d^{2 - 2\epsilon} \bm k_\perp \cdots$. This means that $\mu_0^2$ is the scale for a $\bm k_\LT$ integration.

This argument tells us that in the splitting functions of a $k_\LT$-ordered parton shower, the parton distribution functions should be $\MSbar$ parton distribution functions evaluated at the splitting scale $\mu_0^2$ (or a constant of order 1 times $\mu_0^2$).

Now consider a $\Lambda$-ordered parton shower, with scale variable $\mu_1^2$, or an angle-ordered parton shower, with scale variable $\mu_2^2$. Then the scale variable is related to $\mu_0^2$ by Eq.~(\ref{eq:musubstitution}). To use in this shower, we need modified PDFs $f_{a/A}(x,\mu_\lambda^2)$, which are defined by a new factorization scheme \cite{NSAllOrder} that matches the scheme used to define the parton shower and, in particular, matches the choice of ordering variable. As argued in Ref.~\cite{NSAllOrder}, matching these schemes in such a way that the integrals in $\cU_\cV(\mu_\Lf^2,\mu^2)$ are infrared finite requires that the shower oriented PDFs should obey an evolution equation obtained by making the variable substitution\footnote{Here the treatment of the argument of $\as$ affects the evolution kernel at order $\as^2$ only. We have chosen to include factors of $(1-z)^\lambda$ only in the term with a factor of $1/(1-z)$. This matches our choice in the shower splitting functions.} (\ref{eq:musubstitution}) in the $\MSbar$ evolution equation (\ref{eq:DGLAP0alt}):
\begin{align}
\label{eq:DGLAP1alt}
\mu_\lambda^2 &\frac{d f_{a/A}(x,\mu_\lambda^2)}{d\mu_\lambda^2}
\notag
\\ 
={}& 
\frac{\as(\lambda_\LR \mu_\lambda^2)}{2\pi}\,
\gamma_a\,
f_{a/A}(x,\mu_\lambda^2)\,
\theta(\mu_\lambda^2 > m_0^2(a))
\notag
\\&   
+\int_0^1\!dz\ 
\theta((1-z)^\lambda \mu_\lambda^2 > m_0^2(a))
\notag
\\& \quad\times\!
\Bigg( 
\frac{\as(\lambda_\LR (1-z)^\lambda \mu_\lambda^2)}{2\pi}\,
\frac{2 C_a}{1-z}
\\& \qquad\ \times
\left[
f_{a/A}(x/z,\mu_\lambda^2)
- f_{a/A}(x,\mu_\lambda^2)
\right]
\notag
\\&\qquad
+\frac{\as(\lambda_\LR \mu_\lambda^2)}{2\pi}\,
\sum_{\hat a} \frac{1}{z}\,P_{a\hat a}^{\rm reg}(z)\,
f_{\hat a/A}(x/z,\mu_\lambda^2) 
\Bigg)
\;.
\notag
\end{align}
One could determine the functions $f_{a/A}(x,\mu_\lambda^2)$ from data using the new factorization scheme. That would be a major undertaking, so what we do for \textsc{Deductor} is to use the same functions at a starting scale around 1 GeV as used for the CT14 PDFs, then use Eq.~(\ref{eq:DGLAP1alt}) to determine the PDFs at higher scales. 

Note that at very large scales, the evolution equation (\ref{eq:DGLAP1alt}) is nearly the same as its $\MSbar$ version. However, at smaller values of the scale variable, the two evolution equations are quite different. Thus $f_{a/A}(x,\mu^2)$ for $\mu^2 \gg 1 \GeV^2$ will be substantially different from $f_{a/A}^{\MSbar}(x,\mu^2)$.

\subsection{No-splitting probabilities}
\label{sec:nosplittinglambda}

Consider now the no-splitting probability in a \textsc{Deductor} shower using one of the three ordering choices. As in some of the previous sections, we simplify the notation by assuming that there is only one initial state hadron and one kind of parton. Then the no-splitting probability for ordering type $\lambda$ is a slightly more precise version of Eq.~(\ref{eq:simplePimod}),
\begin{align}
\label{eq:simplePilambda}
\Pi(\mu_{\Lh,\lambda}^2 &, \mu_{s,\lambda}^2, x) 
\notag
\\ ={}& 
\exp\!\bigg(
- \int_{\mu_{\Ls,\lambda}^2}^{\mu_{\Lh,\lambda}^2}\!\frac{d\mu_\lambda^2}{\mu_\lambda^2}
\int\!\frac{dz}{z} \int\!d\phi\
\frac{f(x/z, \mu_\lambda^2)}{f(x,\mu_\lambda^2)}
\notag
\\&\times 
\bigg[
\frac{\as(\lambda_\LR (1-z)^\lambda \mu_\lambda^2)}{2\pi}\, 
G_\lambda^\mathrm{sing}(\mu_\lambda^2, z, \phi)
\\& \quad +
\frac{\as(\lambda_\LR \mu_\lambda^2)}{2\pi}\, 
G_\lambda^\mathrm{reg}(\mu_\lambda^2, z, \phi)
\bigg]
\bigg)
. \notag
\end{align}
Here $\mu_{\Lh,\lambda}^2$ is $c_\lambda$ times the value of the ordering variable for a previous splitting or the ordering variable assigned to the start of the shower, with $c_0 = 1$, $c_1 = 2 p_\La\cdot Q_0/Q_0^2$ and $c_2 = 1$. Similarly, $\mu_{\Ls,\lambda}^2$ is $c_\lambda$ times the value of the ordering variable for a next splitting. The PDFs are the ones appropriate to ordering type $\lambda$. In the exponent, we are integrating over the probability for a potential splitting that did not happen. We divide the splitting function into a part, $G_\lambda^\mathrm{sing}$ that is singular for $(1-z)\to 0$ and a part $G_\lambda^\mathrm{reg}$ that is regular for $(1-z)\to 0$. Then we set the argument of $\as$ for the potential splitting to $\lambda_R (1-z)^\lambda \mu_\lambda^2$ for the singular part and to $\lambda_R \mu_\lambda^2$ for the regular part. One could omit $\lambda_R$ in the regular part, but it seems to us simpler to retain $\lambda_R$ in the argument of $\as$ both terms in Eq.~(\ref{eq:simplePilambda}) and in Eq.~(\ref{eq:DGLAP1alt}). These choices for the argument of $\as$ affect the splitting functions only at order $\as^2$. 

The function $G_\lambda(\mu_\lambda^2, z, \phi)$ is the shower splitting function. We define it by
\begin{equation}
\begin{split}
G_\lambda^\mathrm{sing}(\mu_\lambda^2, z, \phi)
={}& G_0^\mathrm{sing}((1-z)^\lambda\mu_\lambda^2, z, \phi)
\;,
\\
G_\lambda^\mathrm{reg}(\mu_\lambda^2, z, \phi)
={}& G_0^\mathrm{reg}((1-z)^\lambda\mu_\lambda^2, z, \phi)
\;.
\end{split}
\end{equation}
The function $G$ is quite complicated, but we need only two of its features. First, we define $G$ to contain a $\theta$ function that requires\footnote{This becomes $m_0^2(a)$ when we restore flavors $a$ for the splitting partons. Then the shower is ultimately turned off at the scale $m_0^2(a)$ for the light partons, which is normally chosen to be about 1 GeV.}
\begin{equation}
\label{eq:IRcut}
\mu_0^2 > m_0^2
\;.
\end{equation}
This $\theta$ function acts to turn off the parton shower at the scale $m_0^2$. One can then complement the parton shower by a hadronization model to cover physics below this scale. There is also a kinematic limit: $G$ is nonzero only for
\begin{equation}
\label{eq:zlimit0}
\frac{(1-z)^2}{z} > \frac{\mu_0^2}{Q^2}
\;.
\end{equation}
For smaller values of $1 - z$, an initial state splitting is not possible \cite{NSThreshold}. This is directly applicable to a $\bm k_\LT$-ordered shower. This translates to 
\begin{equation}
\label{eq:zlimit1}
\frac{1-z}{z} > \frac{\mu_1^2}{Q^2}
\;,
\end{equation}
which is directly applicable to a $\Lambda$-ordered shower. For an angle-ordered shower, we need
\begin{equation}
\label{eq:zlimit2}
\frac{1}{z} > \frac{\mu_2^2}{Q^2}
\;.
\end{equation}
This is no restriction at all as long as $\mu_2^2 < Q^2$. That is, the angle-ordered shower is not really a hardness-ordered shower because an infinitely soft gluon emission can still have a large angle and thus a large $\mu_2^2$. For an angle-ordered shower, the only restriction on the $z$ integral in Eq.~(\ref{eq:simplePilambda}) comes from the infrared cutoff in Eq.~(\ref{eq:IRcut}),
\begin{equation}
\label{eq:IRcut2}
(1-z)^2 \mu_2^2 > m_0^2
\;.
\end{equation}

\subsection{The PDF property}
\label{sec:PDFpropertylambda}

With this preparation, we can now investigate what happens with the PDF property with the three choices of ordering variables.

For $k_\LT$ ordering ($\lambda = 0$), we have examined $\overline \Pi_\textrm{alt}(t)$, which tests the PDF property, in Fig.~\ref{fig:ductrhoIalt}. Then we examined $\overline \Pi_{\textrm{opt}}(t)$, which includes the operator $\cU_\cV(\mu_\Lf^2,\mu^2)$, in Fig.~\ref{fig:ductrhoIaltV}.

\begin{figure}
\begin{center}

\ifusefigs 

\begin{tikzpicture}
  \begin{axis}[
    title= {\textsc{Deductor}, $\Lambda$ ordered},
    xlabel={$t$}, ylabel={$\overline \Pi_\mathrm{alt}(t)$},
    legend cell align=left,
    every axis legend/.append style = {
      anchor=north east
    },
    xmin=0, xmax=7.0,
    ymin=0, ymax=1.1
    ]
    
    \ratioplot[red,semithick]{fill=red!30!white, opacity=0.5}
    {HardDataLambda.dat}{1}{3}{4}{7}{8}
    \addlegendentry{Hard PDF}
    
    \ratioplot[black,semithick]{fill=black!30!white, opacity=0.5}
    {CTDataLambda.dat}{1}{3}{4}{7}{8}
    \addlegendentry{CT14 PDF}
    
    \ratioplot[blue,semithick]{fill=blue!30!white, opacity=0.5}
    {SoftDataLambda.dat}{1}{3}{4}{7}{8}
    \addlegendentry{Soft PDF}

  \end{axis}
\end{tikzpicture}

\fi

\end{center}
\caption{
Alternative no-splitting fraction, $\overline\Pi_\mathrm{alt}(t)$, versus shower time according to \textsc{Deductor} with $\Lambda$ ordering. This figure is analogous to Fig.~\ref{fig:ductrhoIalt}, which shows $\overline \Pi_\mathrm{alt}(t)$ with $k_\LT$ ordering. The range of $t$ values shown here is different since now $t$ has a different meaning.
}
\label{fig:ductLambdarhoIalt}
\end{figure}
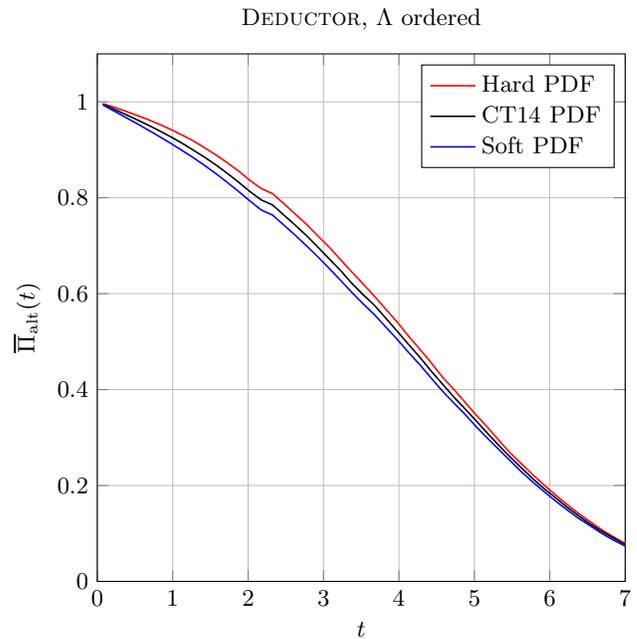

Consider next $\Lambda$ ordering. The comparison for $\overline \Pi_{\textrm{alt}}(t)$ is shown in Fig.~\ref{fig:ductLambdarhoIalt}. There is a significant dependence on the PDF set, but it is smaller than with $k_\LT$ ordering (Fig.~\ref{fig:ductrhoIalt}).

In fact, the operator $\cU_\cV(\mu_\Lf^2,\mu^2)$ is closer to the unit operator for $\Lambda$ ordering than for $k_\LT$ ordering. This is because the restriction (\ref{eq:zlimit1}) for $\Lambda$ ordering is less restrictive than the restriction (\ref{eq:zlimit0}) for $k_\LT$ ordering. The fact that $\cU_\cV(\mu_\Lf^2,\mu^2)$ is closer to the unit operator for $\Lambda$ ordering than for $k_\LT$ ordering reflects the fact that some of the result of summation of threshold logarithms moves from $\cU_\cV(\mu_\Lf^2,\mu^2)$ to the change in PDFs compared to the $\MSbar$ PDFs \cite{NSThreshold}.

\begin{figure}
\begin{center}

\ifusefigs 

\begin{tikzpicture}
  \begin{axis}[
    title= {\textsc{Deductor}, $\Lambda$ ordered},
    xlabel={$t$}, ylabel={$\overline \Pi_{\mathrm{opt}}(t)$},
    legend cell align=left,
    every axis legend/.append style = {
      anchor=north east
    },
    xmin=0, xmax=7.0,
    ymin=0, ymax=1.1
    ]
    
    \ratioplot[red,semithick]{fill=red!30!white, opacity=0.5}
    {HardDataLambda.dat}{1}{5}{6}{7}{8}
    \addlegendentry{Hard PDF}
    
    \ratioplot[black,semithick]{fill=black!30!white, opacity=0.5}
    {CTDataLambda.dat}{1}{5}{6}{7}{8}
    \addlegendentry{CT14 PDF}
    
    \ratioplot[blue,semithick]{fill=blue!30!white, opacity=0.5}
    {SoftDataLambda.dat}{1}{5}{6}{7}{8}
    \addlegendentry{Soft PDF}

  \end{axis}
\end{tikzpicture}

\fi

\end{center}
\caption{
Modified no-splitting fraction $\overline \Pi_{\mathrm{opt}}(t)$, including the threshold operator $\cU_\cV(\mu_\Lf^2,\mu^2)$, versus shower time according to \textsc{Deductor} with $\Lambda$ ordering.
}
\label{fig:ductLambdarhoIaltV}
\end{figure}
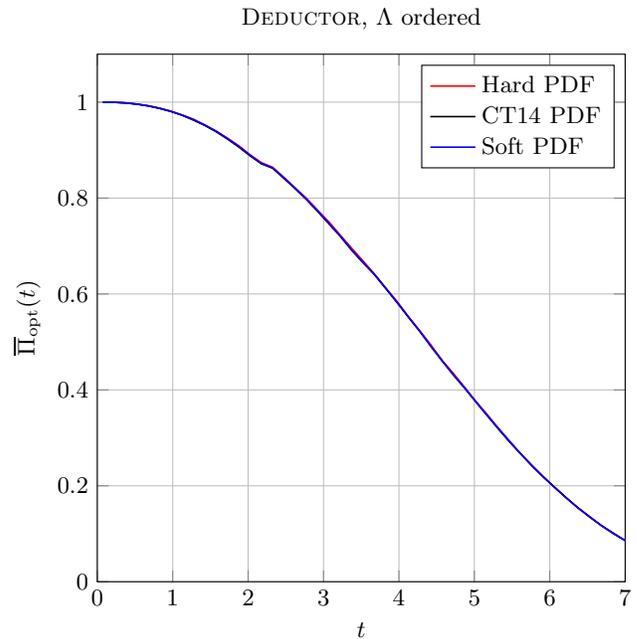

The comparison for $\overline \Pi_{\textrm{opt}}(t)$ is shown in Fig.~\ref{fig:ductLambdarhoIaltV}. The dependence on the PDF set is barely discernible, as it was for $k_\LT$ ordering (Fig.~\ref{fig:ductrhoIaltV}).

\begin{figure}
\begin{center}

\ifusefigs 

\begin{tikzpicture}
  \begin{axis}[
    title= {\textsc{Deductor}, angle ordered},
    xlabel={$t$}, ylabel={$\overline \Pi_\mathrm{alt}(t)$},
    legend cell align=left,
    every axis legend/.append style = {
      anchor=north east
    },
    xmin=0, xmax=3.0,
    ymin=0, ymax=1.1
    ]
    
    \ratioplot[red,semithick]{fill=red!30!white, opacity=0.5}
    {HardDataAngle.dat}{1}{3}{4}{7}{8}
    \addlegendentry{Hard PDF}
    
    \ratioplot[black,semithick]{fill=black!30!white, opacity=0.5}
    {CTDataAngle.dat}{1}{3}{4}{7}{8}
    \addlegendentry{CT14 PDF}
    
    \ratioplot[blue,semithick]{fill=blue!30!white, opacity=0.5}
    {SoftDataAngle.dat}{1}{3}{4}{7}{8}
    \addlegendentry{Soft PDF}

  \end{axis}
\end{tikzpicture}

\fi

\end{center}
\caption{
Alternative no-splitting fraction versus shower time according to \textsc{Deductor} with angle ordering. This figure is analogous to Fig.~\ref{fig:ductrhoIalt}, which shows $\overline \Pi_\mathrm{alt}(t)$ with $k_\LT$ ordering. The range of $t$ values shown here is different since now $t$ has a different meaning.
}
\label{fig:ductAnglerhoIalt}
\end{figure}
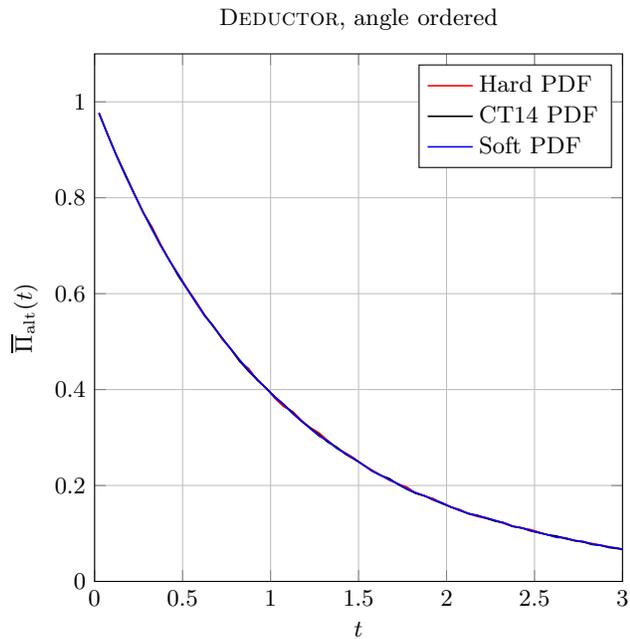

Finally, consider angle ordering.\footnote{We have not used the angle ordering option in \textsc{Deductor} in practical calculations of cross sections. Thus this option is rather untested. Nevertheless, we deem it of interest to test how well the PDF property works with this ordering option.} The restriction (\ref{eq:zlimit2}) for angle ordering provides no restriction at all. Now the upper limit for the $z$ integration in the shower matches the upper limit for the $z$ integration in the evolution of the PDFs. Because of this, we can expect the PDF property for $\overline \Pi_{\textrm{alt}}(t)$ to work to a good approximation. The comparison for $\overline \Pi_{\textrm{alt}}(t)$ is shown in Fig.~\ref{fig:ductAnglerhoIalt}. There are three curves for the three PDF sets, but the differences are small. 

With angle ordering, the operator $\cU_\cV(\mu_\Lf^2,\mu^2)$ is close to the unit operator. In the case examined in this paper, the approximation to $\cU_\cV(\mu_\Lf^2,\mu^2)$ used in \textsc{Deductor} makes $\overline \Pi_{\mathrm{opt}}(t) = \overline\Pi_\mathrm{alt}(t)$ exactly, as explained in Appendix \ref{sec:operators}. Thus the graph for $\overline\Pi_{\mathrm{opt}}(t)$ is the same as Fig.~\ref{fig:ductAnglerhoIalt} and we do not display it. Now all of result of summation of threshold logarithms is contained in the change in PDFs compared to the $\MSbar$ PDFs \cite{NSThreshold}.

\section{Conclusions}
\label{sec:conclusions}

Initial state splittings in backward evolution in a parton shower event generator are characterized by the probability $\overline \Pi(t)$ not to have a splitting between two scales $\mu_\Lh^2$ and $\mu_\Ls^2 = e^{-t}\mu_\Lh^2$. This function is the exponential of an integral of the shower splitting functions times a ratio of PDFs. One can consider another function, $\overline \Pi_\mathrm{alt}(t)$ obtained by multiplying $\overline \Pi(t)$ by the ratio of PDFs with these two scales according to Eq.~(\ref{eq:Pialtdef}). If one were to follow an argument that ignores some complications, one would conclude that the function $\overline \Pi_\mathrm{alt}(t)$ is independent of the parton distribution functions used in the calculation. We call this conclusion the {\em PDF property} of the shower evolution.

One can quite easily test the PDF property numerically by running the parton shower event generator using different PDF sets. We use three PDF sets: CT14, soft, and hard. Each obeys the same evolution equation but the three sets start from different initial conditions. We test \textsc{Pythia} and we test \textsc{Deductor} with $k_\LT$ ordering.

We find that the PDF property does not hold. Thus we conclude that the complications that were ignored in the argument for the PDF property are not just trivial complications but are rather quite significant. It appears that the main difficulty comes from the upper end point of the integral over momentum fraction $z$ in the exponent of $\overline \Pi(t)$. This upper end point was regarded as being close to 1, but in view of the presence of $1/(1-z)$ singularities, it is not close enough.

\textsc{Deductor} has an approximation to another operator, $\cU_\cV(\mu_\Lf^2,\mu^2)$, that adjusts cross sections and has the effect of summing threshold logarithms. Using this operator, one can define another function, $\overline \Pi_\mathrm{opt}(t)$ that should be independent of the PDF set to the accuracy of the approximations to $\cU_\cV(\mu_\Lf^2,\mu^2)$ and $\cN(\mu_\Ls^2,\mu_\Lh^2)$ in \textsc{Deductor}. We find that $\overline \Pi_\mathrm{opt}(t)$ is independent of the PDF set to a good accuracy.

We also test \textsc{Deductor} using its default virtuality based ordering variable $\Lambda^2$ and we test \textsc{Deductor} using angle ordering. In either case, inside the shower one must use shower oriented PDFs that obey a modified evolution equation appropriate to the ordering variable. 

With $\Lambda$ ordering, as expected, the PDF property fails to hold for $\overline \Pi_\mathrm{alt}(t)$ but $\overline \Pi_\mathrm{opt}(t)$ is quite accurately independent of the PDF set. 

With angle ordering, for the case examined in this paper, $\overline \Pi_\mathrm{opt}(t)$ is equal to $\overline \Pi_\mathrm{alt}(t)$. We thus expect $\overline \Pi_\mathrm{alt}(t)$, calculated with PDFs that obey the evolution equation appropriate for angle ordering, to be almost independent of the choice of PDF set. We find that this is the case.

It would be of interest to see the extent to which other parton shower event generators exhibit or do not exhibit the PDF property. A numerically substantial failure to exhibit this property does not indicate that there is something wrong with the probability preserving parton shower. Rather, it says something about the numerical importance of the analogue of the operator $\cU_\cV$. As we see in Eq.~(\ref{eq:sigmaU8}), this operator provides corrections to the hard scattering matrix elements that appear before the probability preserving parton shower starts. A substantial violation of the PDF property indicates that the cross section corrections contained in $\cU_\cV$ are numerically important. The absence of these corrections can be partly alleviated by NLO matching of the hard scattering matrix elements to the shower, but the part of $\cU_\cV$ that is beyond NLO would still be missing. 

In our view, an analogue of $\cU_\cV$ should be included in every parton shower event generator because it is part of a definition of a parton shower from the first principles of QCD perturbation theory \cite{NSAllOrder}. However, most such parton shower programs lack the needed $\cU_\cV$ factor. It is thus of interest that one can, by testing the PDF property, assess the numerical importance of $\cU_\cV$ corrections without having constructed the analogue of $\cU_\cV$ for an existing parton shower event generator.

For the future, it would certainly be desirable to have a parton shower with splitting kernels evaluated at order $\as^2$. This should come along with the operator $\cU_\cV(\mu_\Lf^2,\mu^2)$ defined with an exponent calculated to order $\as^2$. It would then be a good check of the consistency of the initial state evolution in the higher order shower that $\overline \Pi_\mathrm{opt}(t)$ defined in Eq.~(\ref{eq:Pialtmod}) is still independent of the PDF set used in its calculation.

\acknowledgments{ 
This work was supported in part by the United States Department of Energy under grant DE-SC0011640. The work benefited from access to the University of Oregon high performance computer cluster, Talapas.

This project originated from the participation of DS at the 2019 workshop {\em Physics at TeV Colliders} at the \'Ecole de Physique des Houches at Les Houches, France. We thank the Les Houches organizers and we thank, in particular, S.~Prestel for suggesting at the workshop that it would be of interest to test what we have here called the PDF property of parton shower initial state evolution.  
}

\appendix

\section{Operators}
\label{sec:operators}

In this appendix, we briefly explain some of the operators that we use in the main text.

The operators in \textsc{Deductor} are operators on a vector space, the ``statistical space,'' that describes the momenta, flavors, colors, and spins for all of the partons created in a shower as the shower develops. The general theory includes parton spins but \textsc{Deductor} simply averages over spins, so our explanation here will leave out parton spins. With $m$ final state partons, the partons carry labels $\La,\Lb,1,2,\dots,m$, where $\La$ and $\Lb$ are the labels of the two initial state partons. The partons have momenta $\{p\}_m = \{p_\La, p_\Lb, p_1,\dots,p_m\}$ and flavors $\{f\}_m$. We take the partons to be massless: $p_i^2 = 0$. The momenta of initial state partons are specified by their momentum fractions $\eta_\La$ and $\eta_\Lb$: $p_\La = \eta_\La p_\LA$ and $p_\Lb = \eta_\Lb p_\LB$, where $p_\LA$ and $p_\LB$ are the hadron momenta. For color, there are ket color basis states $\ket{\{c\}_m}$ and bra color basis states $\bra{\{c'\}_m}$. We use the trace basis, as described in Ref.~\cite{NSI}. Then the $m$-parton basis states for the statistical space are denoted by $\sket{\{p,f,c',c\}_{m}}$.

Following Ref.~\cite{NSAllOrder}, we begin with an operator $\cU_\mathrm{pert}(\mu_\Ls^2,\mu_\Lh^2)$ that describes evolution from a harder scale $\mu_\Lh^2$ to a softer scale $\mu_\Ls^2$. As explained in the main text, the operator $\cU_\mathrm{pert}(\mu_\Ls^2,\mu_\Lh^2)$ is constructed from Feynman diagrams with approximations that capture their collinear and soft limits. PDFs do not appear in $\cU_\mathrm{pert}(\mu_\Ls^2,\mu_\Lh^2)$. 

We can look at this in a little more detail. $\cU_\mathrm{pert}(\mu_\Ls^2,\mu_\Lh^2)$ is an exponential, as given in Eq.~(74) of Ref.~\cite{NSAllOrder}
\begin{equation}
\label{eq:cUpert}
\cU_{\rm pert}(\mu^2,\mu^{\prime\,2})
= \mathbb{T} \exp\!\left(
\int_{\mu^2}^{\mu^{\prime\,2}}\!\frac{d\mu^2}{\mu^2}\,\cS_{\rm pert}(\mu^2)
\right)
\;,
\end{equation}
where $\mathbb{T}$ indicates $\mu^2$ ordering of the exponential with smaller $\mu^2$ to the left. Working to order $\as$, we have
\begin{equation}
\label{eq:Spertexpansion}
\cS_{\rm pert}(\mu^2)
= \frac{\as(\mu^2)}{2\pi}\,\cS^{(1,0)}_{\rm pert}(\mu^2)
+ \frac{\as(\mu^2)}{2\pi}\,\cS^{(0,1)}_{\rm pert}(\mu^2)
+\cO(\as^2)
,
\end{equation}
The operator $\cS^{(1,0)}_{\rm pert}(\mu^2)$ describes real emissions and contributes to $\cS_\mathrm{split}^\mathrm{pert}(\mu^2)$ in Eq.~(\ref{eq:Upertevolution}) in the main text. The operator $\cS^{(0,1)}_{\rm pert}(\mu^2)$ represents virtual exchange graphs and contributes to $\cS_\mathrm{no-split}^\mathrm{pert}(\mu^2)$ in Eq.~(\ref{eq:Upertevolution}).

The operator $\cS^{(1,0)}_{\rm pert}(\mu^2)$ is defined in Eq.~(6.5) of Ref.~\cite{NSThreshold} as the operator $\cS^{(1,0)}(\mu^2)$ with the PDF factors removed, where $\cS^{(1,0)}(\mu^2)$ is the operator called $\cH_I(\mu^2)$ given in Eq.~(5.7) of Ref.~\cite{NScolor} (which is based on Eq.~(8.26) of Ref.~\cite{NSI}). 
When $\cS^{(1,0)}(\mu^2)$ is applied to a statistical basis state $\sket{\{p,f,c',c\}_{m}}$ with $m$ partons, it produces a linear combination of basis states $\sket{\{\hat p,\hat f,\hat c',\hat c\}_{m+1}}$ with $m+1$ partons by adding a new parton with label $m+1$. The operator $\cS^{(1,0)}(\mu^2)$ is defined by
\begin{equation}
\begin{split}
\label{eq:HIdef}
\big(\{\hat p{}&,\hat f,\hat c',\hat c\}_{m+1}\big|
\cS^{(1,0)}(\mu^2)\sket{\{p,f,c',c\}_{m}}
\\={}&
\sum_{l,k}
\delta(\mu^2 - \mu^2_l(\{\hat p,\hat f\}_{m+1}))
\\&\times
(m+1)
\sbra{\{\hat p,\hat f\}_{m+1}}{\cal P}_{l}\sket{\{p,f\}_m}
\\&\times
\frac
{n_\Lc(a) n_\Lc(b)\,\eta_{\La}\eta_{\Lb}}
{n_\Lc(\hat a) n_\Lc(\hat b)\,
 \hat \eta_{\La}\hat \eta_{\Lb}}\,
\frac{
f_{\hat a/A}(\hat \eta_{\La},\mu^{2}_{F})
f_{\hat b/B}(\hat \eta_{\Lb},\mu^{2}_{F})}
{f_{a/A}(\eta_{\La},\mu^{2}_{F})
f_{b/B}(\eta_{\Lb},\mu^{2}_{F})}
\\&\times \frac{1}{2}\bigg[
\theta(k = l)\,
\theta(\hat f_{m+1} \ne g)\,
\overline w_{ll}(\{\hat p,\hat f\}_{m+1})
\\&\quad +
\theta(k = l)\,
\theta(\hat f_{m+1} = g)
\\&\qquad\times
[\overline w_{ll}(\{\hat p,\hat f\}_{m+1})
- \overline w_{ll}^{\rm eikonal}(\{\hat p,\hat f\}_{m+1})]
\\&\quad -
\theta(k\ne l)\,
\theta(\hat f_{m+1} = g)
\\&\qquad\times
A'_{lk}(\{\hat p\}_{m+1})\overline w_{lk}^{\rm dipole}(\{\hat p,\hat f\}_{m+1})
\bigg]
\\&\hskip 0.2 cm \times
\bigg[
\sbra{\{\hat c',\hat c\}_{m+1}}
t^\dagger_l(f_l \to \hat f_l + \hat f_{m+1})
\\&\qquad\otimes 
t_k(f_k \to \hat f_k + \hat f_{m+1})
\sket{\{c',c\}_m}
\\&\quad
+
\sbra{\{\hat c',\hat c\}_{m+1}}
t^\dagger_k(f_k \to \hat f_k + \hat f_{m+1}) 
\\&\qquad\otimes 
t_l(f_l \to \hat f_l + \hat f_{m+1})
\sket{\{c',c\}_m}
\bigg]\;\;.
\end{split}
\end{equation}

In the first line on the right-hand side of Eq.~(\ref{eq:HIdef}), we have a sum over parton indices $l$ and $k$. The parton with label $l$ is the one that splits. There is another label $k$ so that we can include graphs that represent quantum interference between emission of a gluon from parton $l$ and from another parton $k$. We call parton $k$ the helper parton. For the quantum interference terms, we have $k \ne l$. There are also graphs that do not represent quantum interference. For these, $k = l$. 

The first line on the right-hand side of Eq.~(\ref{eq:HIdef}) also includes a delta function that specifies the definition of the shower ordering variable $\mu^2$, using a function $\mu^2_l(\{\hat p,\hat f\}_{m+1})$. To understand this, we need to define the important kinematic variables in a splitting, as given in Appendix A of Ref.~\cite{NSThreshold}. It is convenient to use a dimensionless virtuality variable $y$ defined, for the splitting of parton $l$ (either an initial state or final state parton) by
\begin{equation}
y = \frac{2 \hat p_l\cdot \hat p_{m+1}}{2 p_l\cdot Q}
\;,
\end{equation}
where $Q = p_\La + p_\Lb$ is the total momentum of the final state partons in the state $\sket{\{p,f,c',c\}_{m}}$. For $l = \La$, we note that $2 p_l\cdot Q = Q^2$. The default ordering variable in \textsc{Deductor} is 
\begin{equation}
\Lambda^2 = y Q_0^2
\;,
\end{equation}
where $Q_0$ is the total momentum of the final state partons at the start of the shower. Thus for a $\Lambda$ ordered shower we can use the scale variable
\begin{equation}
\mu^2_l(\{\hat p,\hat f\}_{m+1}) = y Q^2
\end{equation}
to describe what is fixed for a splitting starting with an $m$-parton state with total momentum $Q$.

We will be particularly interested in the splitting of an initial state parton, say parton ``a.'' The parton has a momentum fraction $\eta_\La$ before the splitting and momentum fraction $\hat \eta_\La$ after the splitting (``after'' in the sense of backward evolution). We define the momentum fraction variable $z$ of this splitting by
\begin{equation}
z = \frac{\eta_\La}{\hat\eta_\La}
\;.
\end{equation}
The part of $\hat p_{m+1}$ orthogonal to the momenta of both incoming partons after the splitting, $\hat p_\La$ and $\hat p_\Lb$, is $\hat {\bm p}_{m+1}^\perp$, whose square is
\begin{equation}
(\hat {\bm p}_{m+1}^\perp)^2  = y (1 - z - zy)\, Q^2
\;.
\end{equation}
This quantity cannot be negative. Thus $(1 - z - zy) > 0$, or
\begin{equation}
z < \frac{1}{1+y}
\;.
\end{equation}
This inequality is very important to the analysis in this paper. It imposes an upper limit on $z$. If $y \ll 1$, the limit is very close to $z = 1$. However, the size of $(1-z)$ can be important because the splitting kernel is large near $z = 1$.

We define a transverse momentum variable $\bm k_\perp^2$ for an initial state splitting. The quantity $(\hat {\bm p}_{m+1}^\perp)^2$ vanishes when the emitted parton is collinear to parton ``a,'' which corresponds to $y \to 0$ with fixed $(1-z)$. However, it also vanishes when the emitted parton is collinear to parton ``b,'' which corresponds to $(1 - z - zy) \to 0$ with fixed $y$. We prefer a variable that matches $(\hat {\bm p}_{m+1}^\perp)^2$ in the collinear limit, but does not vanish in the anticollinear limit $(1 - z - zy) \to 0$. We thus define
\begin{equation}
\label{eq:kperpsqdef}
\bm k_\LT^2  = y (1 - z)\, Q^2
\;.
\end{equation}
We use this as the ordering variable for a $k_\LT$-ordered shower in \textsc{Deductor}. Thus for a $k_\LT$-ordered shower we take
\begin{equation}
\mu^2_l(\{\hat p,\hat f\}_{m+1}) = y (1 - z)\, Q^2
\end{equation}
in the first line on the right-hand side of Eq.~(\ref{eq:HIdef}) when $l = \La$.

The second line on the right-hand side of Eq.~(\ref{eq:HIdef}) contains the function that defines the momentum mapping. For an initial state splitting of parton ``a,'' this mapping supplies the momentum $p_\La - \hat p_\La + \hat p_{m+1}$ needed for the splitting. \textsc{Deductor} uses a global momentum mapping as specified in Eq.~(A.16) of Ref.~\cite{ShowerTime}.

In the third line, we have a ratio of parton distribution functions and associated initial state factors. For a final state splitting, this ratio is 1. For an initial state splitting, this ratio replaces the parton distribution functions at the previous momentum fraction $\eta_\La$ or $\eta_\Lb$ by the parton distribution function at the new momentum fraction after the splitting. If we want $\cS_\mathrm{pert}^{(1,0)}(\mu^2)$ instead of $\cS^{(1,0)}(\mu^2)$, we simply delete this line.

Skipping for the moment to the last four lines of Eq.~(\ref{eq:HIdef}), we see the factor that relates the new color state $\{\hat c',\hat c\}_{m+1}$ to the old color state $\{c',c\}_{m}$. We simply insert the proper color matrices for the emission of a gluon from parton $l$ and from parton $k$ or for the splitting of a gluon into a $q \bar q$ state. This factor is described in Ref.~\cite{NScolor}. This factor is given in Eq.~(\ref{eq:HIdef}) for exact color. Normally, \textsc{Deductor} uses the LC+ approximation, which is described in some detail in Ref.~\cite{NScolor}.

This brings us to the remaining lines in Eq.~(\ref{eq:HIdef}), which contain the splitting functions. The function $w_{lk}^{\rm dipole}(\{\hat p,\hat f\}_{m+1})$ is the familiar eikonal function that describes interference of the emission of a gluon from parton $l$ with the emission of a gluon from parton $k$ in Feynman gauge. It is given in Eq.~(5.3) of Ref.~\cite{NScolor}:
\begin{equation}
\overline w_{lk}^{\,\rm dipole}
= 4\pi\as\
\frac{
2\hat p_{k}\cdot \hat p_l}
{\hat p_{m+1}\cdot \hat p_k\ \hat p_{m+1}\cdot \hat p_l}
\;.
\end{equation}
This function is multiplied by a function $A'_{lk}(\{\hat p\}_{m+1})$ that serves to partition the $l$-$k$ dipole emission into a part considered to be an emission from parton $l$ and a part considered to be an emission from parton $k$. (See Eq.~(5.8) of Ref.~\cite{NScolor}.) The choice in \textsc{Deductor} for $A'_{lk}$ is given in Eq.~(7.12) of Ref.~\cite{NSspin}:
\begin{equation}
A'_{l k}(\{\hat p\}_{m+1}) = \frac{\hat p_{m+1}\cdot \hat p_k\ \hat p_l\cdot \hat Q}
{\hat p_{m+1}\cdot \hat p_k\ \hat p_l\cdot \hat Q
+ \hat p_{m+1}\cdot \hat p_l\ \hat p_k\cdot \hat Q}
\;.
\end{equation}
where $\hat Q = \hat p_\La + \hat p_\Lb$.

The functions $\overline w_{ll}(\{\hat p,\hat f\}_{m+1})$ describe splittings of parton $l$ in both the ket and bra states, with the use of a physical gauge. For instance, for emission of a gluon from initial state quark line ``a,'' the definition is (from Eq.~(2.26) of Ref.~\cite{NSII})
\begin{equation}
\begin{split}
\overline w_{\La\La} ={}& \frac{4\pi\as}{2(p_\La\cdot p_\Lb)^2}\,
\frac{1}{(2\,\hat p_\La\cdot \hat p_{m+1})^2}\
D_{\mu\nu}(\hat p_{m+1},\hat Q)
\\&\times
\frac{1}{4}{\rm Tr}\left[
\s{\hat p}_\La \gamma^\mu 
(\s{\hat p}_\La - \s{\hat p}_{m+1})
\s{p}_\Lb
\s{p}_\La \s{p}_\Lb 
(\s{\hat p}_\La - \s{\hat p}_{m+1})
\gamma^\nu
\right]
\;,
\end{split}
\end{equation}
where
\begin{equation}
\begin{split}
D_{\mu\nu}(\hat p_{m+1},\hat Q) ={}&
-g^{\mu\nu} 
+ \frac{\hat p^\mu_{m+1} \hat Q^\nu + \hat Q^\mu\hat p^\nu_{m+1} }
{p^\mu_{m+1} \cdot \hat Q}
\\ &
- \frac{\hat Q^2\,\hat p^\mu_{m+1} \hat p^\nu_{m+1} }
{(p^\mu_{m+1} \cdot \hat Q)^2}
\;.
\end{split}
\end{equation}
Using Eq.~(2.38) of Ref.~\cite{NSII}), this is\footnote{Ref.~\cite{NSII} uses variables $y'$, $z'$, and $x'$, denoted as $y$, $z$, and $x$, with $y' = zy$, $z' = [1-z(1+y)]/[1-z y]$, and $x' = 1-z$, so one has to translate the results to the current notation. Also, $\overline w_{ll}$ is denoted by $\overline W_{ll}$ in Ref.~\cite{NSII}).}
\begin{equation}
\overline w_{\La\La} =
\frac{4\pi\as}{p_\La\cdot Q}\,\frac{1}{yz}
\left\{\frac{1+z^2}{1-z}
- y z^2\left[\frac{2}{(1-z)^2}+ 1\right]
\right\}
\;.
\end{equation}
The quantity in braces here is approximately the coefficient of $C_\LF$ in the splitting kernel $\widehat P(z)$ for the evolution of the quark PDF, $[1+z^2]/(1-z)$, as long as $y \ll (1-z)$. However, it behaves quite differently when $y \sim (1-z)$.

The remaining functions $\overline w_{ll}$ are given in Ref.~\cite{NSII}.

Also in Eq.~(\ref{eq:HIdef}) we have the function $\overline w_{ll}^{\rm eikonal}(\{\hat p,\hat f\}_{m+1})$. This function gives the eikonal approximation to gluon emission in the physical gauge that we used for $\overline w_{ll}(\{\hat p,\hat f\}_{m+1})$. It is given in Eq.~(2.10) of Ref.~\cite{NSII}):
\begin{equation}
\overline w_{ll}^{\rm eikonal}(\{\hat p,\hat f\}_{m+1})
= 4\pi\as\,
\frac{\hat p_l\cdot D(\hat p_{m+1},\hat Q) \cdot \hat p_l}
{(p_{m+1}\cdot \hat p_l)^2}
\;.
\end{equation}
For an initial state splitting of parton ``a,'' this function is given in Eq.~(2.39) of Ref.~\cite{NSII}:
\begin{equation}
\overline w_{\La\La}^{\rm eikonal} =
\frac{4\pi\as}{p_\La\cdot Q}\,\frac{2}{y}\,
\frac{1-z(1+y)}{(1-z)^2}
\;.
\end{equation}
The difference between $\overline w_{\La\La}$ and $\overline w_{\La\La}^{\rm eikonal}$ for gluon emission from an initial state quark is
\begin{equation}
\overline w_{\La\La} - \overline w_{\La\La}^{\rm eikonal} =
\frac{4\pi\as}{p_\La\cdot Q}\,\frac{1}{yz}
\left\{1-z - z^2 y
\right\}
\;.
\end{equation}
We see that the subtraction removes the $\textrm{soft}\times \textrm{collinear}$ singularity at $y \to 0$, $(1-z) \to 0$. This singularity is still present, but is contained in the terms in Eq.~(\ref{eq:HIdef}) with $k\ne l$.

This completes the description of $\cS^{(1,0)}_\mathrm{pert}(\mu^2)$, which contributes to Eq.~(\ref{eq:cUpert}) in the exponent of $\cU_{\rm pert}(\mu^2,\mu^{\prime\,2})$ at order $\as$.

We also need $\cS^{(0,1)}_{\rm pert}(\mu^2)$, which is derived from virtual graphs and has a real part and an imaginary part:
\begin{equation}
\cS^{(0,1)}_\mathrm{pert}(\mu^2) =
\cS^{(0,1)}_\mathrm{Re}(\mu^2) + \cS^{(0,1)}_{\mi\pi}(\mu^2)
\;.
\end{equation}
The imaginary part, which is proportional to $\mi \pi$, can be found in Eq.~(A41) of Ref.~\cite{NSThresholdII}. It is very simple but is not of much interest for this paper. The real part consists of one contribution for each parton,
\begin{equation}
\cS^{(0,1)}_\mathrm{Re}(\mu^2) = \sum_l \cS^{(0,1)}_{\mathrm{Re},l}(\mu^2)
\;.
\end{equation}
The contribution from the initial state parton ``a'' is given in Eqs.~(A38) and (A34) of Ref.~\cite{NSThresholdII}, omitting the $\mi\pi$ term:\footnote{The operator that we here call $[\as/(2\pi)]\,\cS^{(0,1)}_{\mathrm{Re},l}$ is denoted by the real part of $-\cS_l^\mathrm{pert}$ in Ref.~\cite{NSThresholdII}.}
\begin{equation}
\begin{split}
\label{eq:S1}
\cS^{(0,1)}_{\mathrm{Re},\La}(\mu^2)&\sket{\{p,f,c',c\}_{m}} 
\\={}& 
\bigg\{\sum_{k \ne \La,\Lb}
\int_{z_k}^{1/(1+y)}\!dz
\\&\quad\times
\left[
\frac{1}{\sqrt{(1-z)^2 +  y^2/\psi_{\La k}^2}} - \frac{1}{1-z}
\right]
\\& \quad\times
\big(
[(\bm{T}_\La\cdot \bm{T}_k)\otimes 1]
+ [1 \otimes (\bm{T}_\La\cdot \bm{T}_k)]
\big)
\\&
+ \left[\gamma_\La + 2 C_\La \log(y)\right]
[1\otimes 1]
\bigg\}
\\&\times
\sket{\{p,f,c',c\}_{m}}
\;.
\end{split}
\end{equation}
This result uses \textsc{Deductor}'s default $\Lambda$ ordering to define $\mu^2 = \mu_1^2 = y Q^2$ according to Eq.~(\ref{eq:LambdaDef}) and uses the approximation that $y \ll 1$, but not $y \ll \psi_{ak}$. Each term corresponds to exchanging a gluon between lines ``a'' and $k$, with corresponding color operators $\bm{T}_\La\cdot \bm{T}_k$ applied either to the ket color state $\ket{\{c\}_m}$ or the bra color state $\bra{\{c'\}_m}$. Thus, in general, a gluon exchange changes the color state. However, if we use the LC+ approximation, these operators simply multiply the state by an eigenvalue when partons $k$ and ``a'' are color connected and give zero otherwise. The parameter $\psi_{\La k}$ is
\begin{equation}
\label{eq:psiak}
\psi_{ak} = \frac{1 -  \cos\theta_{ak}}{\sqrt{8(1 +  \cos\theta_{ak})}}
\;,
\end{equation}
where $\theta_{\La k}$ is the angle between the two partons in the reference frame in which $\vec Q = 0$. The parameter $z_k$ is defined by
\begin{equation}
p_k \cdot p_\Lb = (1-z_k) p_\La \cdot p_\Lb
\;.
\end{equation}
The constants $C_\La$ and $\gamma_\La$ are defined in Eqs.~(\ref{Pregdef}) and (\ref{eq:gammaf}). This completes the definition of the operator $S_{\rm pert}(\mu^2)$ at first order in $\as$. 

We also need the operator $\cS(\mu^2)$ that generates the shower according to
\begin{equation}
\label{eq:cU}
\cU(\mu_\Ls^2,\mu_\Lh^2)
= \mathbb{T} \exp\!\left(
\int_{\mu_\Ls^2}^{\mu_\Lh^2}\!\frac{d\mu^2}{\mu^2}\,\cS(\mu^2)
\right)
\;,
\end{equation}
At first order, this operator has the form given in Eq.~(125) of Ref.~\cite{NSAllOrder},
\begin{equation}
\begin{split}
\label{eq:S1storder2}
\cS(\mu^2) ={}& \frac{\as(\mu^2)}{2\pi}\,
\cF(\mu^2)\,
\cS_{\rm pert}^{(1,0)}(\mu^2)\,
\cF^{-1}(\mu^{2})
\\&
- \frac{\as(\mu^2)}{2\pi}\,
[\cF(\mu^2)\circ\bar{\cS}^{(1,0)}(\mu^2)]\,\cF^{-1}(\mu^2)
\\&
+ \frac{\as(\mu^2)}{2\pi}\,\cS_{\mi \pi}^{(0,1)}(\mu^2)
+ \cO(\as^2)
\;.
\end{split}
\end{equation}
We have already met the operator $\cS_{\rm pert}^{(1,0)}(\mu^2)$ in Eq.~(\ref{eq:HIdef}). The operators $\cF(\mu^2)$ and $\cF^{-1}(\mu^{2})$ insert the PDF factor in the third line on the right-hand side of Eq.~(\ref{eq:HIdef}), turning $\cS_{\rm pert}^{(1,0)}(\mu^2)$ into 
\begin{equation}
\cS^{(1,0)}(\mu^2) = \cF(\mu^2)\,
\cS_{\rm pert}^{(1,0)}(\mu^2)\,
\cF^{-1}(\mu^{2})
\;.
\end{equation}
The operator $\cS_{\mi \pi}^{(0,1)}(\mu^2)$ inserts phase factors and is not important for the present paper.

We have next an operator $[\cF(\mu^2) \circ\bar{\cS}^{(1,0)}(\mu^2)]\,\cF^{-1}(\mu^2)$, where the $\circ$ denotes a convolution in momentum fraction in the notation of Ref.~\cite{NSAllOrder}. For instance, the first order evolution equation (\ref{eq:DGLAP}) for the parton distribution functions is written as
\begin{equation}
\label{eq:PDFevolution}
\mu^2 \frac{d}{d\mu^2}\, \cF(\mu^2)
= \frac{\as(\mu^2)}{2\pi}\, [\cF(\mu^2) \circ \cP^{(1)}(\mu^2)]
\;.
\end{equation}

The operator $[\cF(\mu^2) \circ\bar{\cS}^{(1,0)}(\mu^2)]\,\cF^{-1}(\mu^2)$ is easy to understand. When evaluated in the LC+ approximation, this operator is a very familiar part of a parton shower. When multiplied by $\as/(2\pi)$ and applied to a state $\sket{\{p,f,c',c\}_{m}}$, it gives an eigenvalue equal to the total probability for one of the partons in this state to split at scale $\mu^2$. Then
\begin{equation}
\begin{split}
\Pi ={}&
\exp\!\bigg(
-\int_{\mu_\Ls^2}^{\mu_\Lh^2}\!\frac{d\mu^2}{\mu^2}\,
\frac{\as(\mu^2)}{2\pi}
\\&\qquad\times
[\cF(\mu^2)\circ\bar{\cS}^{(1,0)}(\mu^2)]\,\cF^{-1}(\mu^2)
\bigg)
\end{split}
\end{equation}
is the Sudakov factor representing the probability not to have a splitting between scales $\mu_\Lh^2$ and $\mu_\Ls^2$. When evaluated using full color, $[\cF(\mu^2) \circ\bar{\cS}^{(1,0)} (\mu^2)]\,\cF^{-1}(\mu^2)$ is defined by integrating ${\cS}^{(1,0)} (\mu^2)$ over the splitting variables and rearranging the color operators. It is evaluated in some detail in Appendix B of Ref.~\cite{NSThreshold}, where it is denoted by $\cV(\mu^2)$.

When we use the shower generator $\cS(\mu^2)$ to define $\cU(\mu_\Ls^2,\mu_\Lh^2)$ according to Eq.~(\ref{eq:cU}), the construction in Eq.~(\ref{eq:S1storder2}) guarantees that the shower preserves the total probability: the sum of the cross sections for the system to be in any of the possible states at the end of the shower equals the sum of the cross sections for the system to be in any state at the start of the shower. 

Now we turn to the operator $\cU_\cV(\mu_\Ls^2,\mu_\Lh^2)$, which is generated by the operator $\cS_\cV(\mu^2)$ according to Eq.~(\ref{eq:UVexponential}). The relation (\ref{eq:Udef}) between $\cU(\mu_\Ls^2,\mu_\Lh^2)$ and  $\cU_\mathrm{pert}(\mu_\Ls^2,\mu_\Lh^2)$ tells us what $\cS_\cV(\mu^2)$ has to be. At first order in $\as$, we have, according to Eq.~(124) of Ref.~\cite{NSAllOrder},
\begin{equation}
\begin{split}
\label{eq:SV1storder}
\cS_\cV(\mu^2) ={}&
\frac{\as(\mu^2)}{2\pi}\,[\cF(\mu^2)\circ\bar{\cS}^{(1,0)}(\mu^2)]\,\cF^{-1}(\mu^2) 
\\&
+ \frac{\as(\mu^2)}{2\pi}\,{\cS}_{\rm Re}^{(0,1)}(\mu^2)
\\&
- \frac{\as(\mu^2)}{2\pi}\, [\cF(\mu^2)\circ \cP^{(1)}(\mu^2)]\,
\cF^{-1}(\mu^2)
\\&
+ \cO(\as^2)
\;.
\end{split}
\end{equation}
We have introduced the first two terms in this result. In the third term, the parton distribution functions are convolved with the first order PDF evolution kernel $\cP^{(1)}(\mu^2)$, as in Eq.~(\ref{eq:PDFevolution}).

In Ref.~\cite{NSThresholdII}, the first order part of $\cS_\cV(\mu^2)$ is denoted by $\cV(\mu^2) - \cS(\mu^2)$ except that we drop the $\mi\pi$ term in $\cS(\mu^2)$. There are contributions associated with each parton $l$ on which $\cS_\cV(\mu^2)$ acts:
\begin{equation}
\cS_\cV(\mu^2) = \cS_{\cV,\La}(\mu^2) 
+ \cS_{\cV,\Lb}(\mu^2)
+ \sum_{l>0} \cS_{\cV,l}(\mu^2)
\;.
\end{equation}
For final state partons, $l > 0$, the net contribution at first order vanishes because of real-virtual cancellations. For the two initial state partons, there are cancellations, but there are some terms left over that are not suppressed by a power of $y$. These terms are given for parton ``a'' in Eq.~(A55) of Ref.~\cite{NSThresholdII}.\footnote{In Eq.~(A55) of Ref.~\cite{NSThresholdII}, the sum over $k$ in the third term included $k=\Lb$, but this contribution is exactly zero, so we have written the sum in Eq.~(\ref{eq:VS5alt}) to exclude $k=\Lb$. Additionally, we have found that the integral $I_k$ is small, so the fourth term in Eq.~(\ref{eq:VS5alt}) is not included in the current version of \textsc{Deductor}.} We state the result for a general state $\sket{\{p,f,c',c\}_{m}}$. In the special case examined numerically in this paper, there are two initial state partons and no final state colored partons. Then the contributions in the general formula that come from final state partons $k \ne \La,\Lb$ are absent. The general formula is

\begin{widetext} 

\begin{equation}
\begin{split}
\label{eq:VS5alt}
\cS_{\cV,\La}(\mu^2)&\sket{\{p,f,c',c\}_{m}} 
\\
={}& 
\Bigg\{
\int_{1/(1+y)}^{1}\!\frac{dz}{z}\ 
\frac{\as((1-z)\lambda_\LR y Q^2)}{2\pi}\,
\theta((1-z) y Q^2 > m_0^2(a))\,
\bigg(
1
-\frac{f_{a/A}(\eta_{\La}/z, y Q^2)}
{f_{a/A}(\eta_{\La}, y Q^2)}
\bigg)
\frac{2 zC_a}{1-z}\,
[1\otimes 1]
\\&  
- \sum_{\hat a} 
\int_{1/(1+y)}^1\!\frac{dz}{z}\ 
\frac{\as(\lambda_\LR y Q^2)}{2\pi}\,
\theta((1-z) y Q^2 > m_0^2(a))\,
P_{a\hat a}^{\rm reg}\!\left(z\right)
\,
\frac{f_{\hat a/A}(\eta_{\La}/z, y Q^2)}
{f_{a/A}(\eta_{\La}, y Q^2)}\,
[1\otimes 1]
\\&  
-\int_0^{1/(1+y)}\!\frac{dz}{z}\ 
\frac{\as((1-z)\lambda_\LR y Q^2)}{2\pi}\,
\theta((1-z) y Q^2 > m_0^2(a))\,
\left(
1
-\frac{
f_{a/A}(\eta_{\La}/z, y Q^2)}
{f_{a/A}(\eta_{\La}, y Q^2)}
\right)
\\&\quad\times
\sum_{k\ne \La, \Lb}\,
\left(
\frac{z}{1-z}
-z\,v(y,z,\theta_{\La k})
\right)\,
\big([(\bm T_\La\cdot \bm T_k)\otimes 1] + [1 \otimes (\bm T_\La\cdot \bm T_k)]\big)
\\&  
+ 
\sum_{k \ne \La,\Lb}
I_k(y,\xi_k,z_k)\,
\big(
[(\bm{T}_\La\cdot \bm{T}_k)\otimes 1]
+ [1 \otimes (\bm{T}_\La\cdot \bm{T}_k)]
\big)
\Bigg\}
\sket{\{p,f,c',c\}_{m}}
+\cO(\as^2)
\;.
\end{split}
\end{equation}
Here $v(y,z,\theta_{\La k})$ is defined in Eq.~(A44) of Ref.~\cite{NSThresholdII}:
\begin{equation}
\label{eq:vresult}
v(y,z,\theta_{\La k}) = 
\frac{z}{1-z}\,\frac{1+y}{1+zy}\ 
\frac{1-\delta}{\sqrt{(1-\delta)^2 + 4 x^2 \delta}}
+ \frac{1}{1 + z y} 
\;,
\end{equation}
with
$
x = {zy}/({1-z})
$
and
$
\delta = (1+zy)\left(1+y\right)(1+\cos\theta_{\La k})/2
$. 
We note that $x$ runs from 0 to 1 when $z$ ranges from 0 to its upper limit, $1/(1+y)$, and that $\delta > 0$. However, $\delta$ can be larger than 1 when $\theta_{\La k}$ is small. We have also defined the integral
\begin{equation}
\begin{split}
I_k(y,\xi_k,z_k)
={}&
\int_0^{1/(1+y)}\!dz\
\frac{\as((1-z)\lambda_\LR y Q^2)}{2\pi}\,
\theta((1-z) y Q^2 > m_0^2(a))
\\&\times
\left[
\frac{\theta(z > z_k)}{\sqrt{(1-z)^2 +  y^2/\psi_{\La k}^2}} 
- \frac{\theta(z > z_k)}{1-z}
- v(y,z,\theta_{\La k})
\right]
\;.
\end{split}
\end{equation}

\end{widetext} 

In Eq.~(\ref{eq:VS5alt}), we have divided the PDF evolution kernel $\widehat P_{a\hat a}(z)$ into a part with a $1/(1-z)$ singularity and a nonsingular part $P_{a\hat a}^{\rm reg}\!\left(z\right)$ according to Eq.~(\ref{Pregdef}). In the coefficient of $P_{a\hat a}^{\rm reg}\!\left(z\right)$, we set the argument of $\as$ to  $\lambda_\LR y Q^2$, where $\lambda_\LR$ is defined in Eq.~(\ref{eq:CMW}). In all of the other terms, we set the argument of $\as$ to  $(1-z)\lambda_\LR y Q^2$, using our definition (\ref{eq:kperpsqdef}) of $\bm k_\LT^2 = (1-z) y Q^2$. These choices could be regarded as somewhat arbitrary, but they affect the operator $\cS_{\cV}(\mu^2)$ only at order $\as^2$. In all of the terms in Eq.~(\ref{eq:VS5alt}), we include a theta function that restricts $\bm k_\LT^2 = (1-z) y Q^2$ to be greater than $m_0^2(a)$, as in Eq.~(\ref{eq:DGLAP0alt}), where $m_0^2(a)$ is the greater of a shower end scale of order $1 \GeV^2$ and the mass of parton with flavor $a$.

The result for $\cS_{\cV,\La}(\mu^2)$ in Eq.~(\ref{eq:VS5alt}) is not simple. However, the main contribution comes from the first term. This term is simple. It results from a near cancellation of two pieces. One is a contribution from the shower, in which $z$ is integrated in the range $0 < z < 1/(1+y)$. The second is a contribution from PDF evolution, in which $z$ is integrated in the range $0 < z < 1$. We are left with an integration over the range $1/(1+y) < z < 1$. For $y \ll 1$, this is a tiny range. However, this term contains a factor $1/(1-z)$, which is large in this range and is singular in the limit $z \to 1$. There is no actual singularity because the PDF factor vanishes when $z \to 1$. Nevertheless, the PDFs are fast varying near $z = 1$ if $\eta_\La$ is large. Thus this term can be substantial, as we have seen in the numerical results in this paper.

The results in this Appendix are for $\Lambda$ ordering. As explained in the main text, if we want $k_\LT$ ordering, we should replace $y$ by $y_0 = (1-z) y$ as the variable held constant in the $z$ integration. Then the integration range $z > 1/(1+y)$ becomes $(1-z)^2/z < y_0$. If we want angle ordering, we should replace $y$ by $y_2 = y/(1-z)$ as the variable held constant. Then the integration range $1/(1+y) < z < 1$ becomes $1/y_2<z<1$. As long as the angle variable $y_2$ is smaller than 1, the integration range vanishes. Thus the first two terms in Eq.~(\ref{eq:VS5alt}) give zero for angle ordering. The third and fourth terms are not present in the case examined in this paper, in which the only helper parton index is $k=\Lb$.




\begin{thebibliography}{99}


\bibitem{sjostrand}
  T.~Sjostrand,
  {\em A Model for Initial State Parton Showers},
   \href{http://dx.doi.org/10.1016/0370-2693(85)90674-4}
  {Phys.\ Lett.\  {\bf 157B}, 321 (1985)}.
  [\href{http://inspirehep.net/search?p=doi:10.1016/0370-2693(85)90674-4}
  {\textsc{inSPIRE}}].

\bibitem{pythia}
  T.~Sj\"ostrand {\it et al.},
  {\em An Introduction to PYTHIA 8.2},
  \href{http://dx.doi.org/10.1016/j.cpc.2015.01.024}
  {Comput.\ Phys.\ Commun.\  {\bf 191}, 159 (2015)}
  [\href{http://inspirehep.net/search?p=doi:10.1016/j.cpc.2015.01.024}
  {\textsc{inSPIRE}}].

\bibitem{Deductor}
  Z.~Nagy and D.~E.~Soper,
  {\em A parton shower based on factorization of the quantum density matrix},
  \href{http://dx.doi.org/10.1007/JHEP06(2014)097}
  {JHEP {\bf 1406}, 097 (2014)}
  [\href{http://inspirehep.net/search?p=find+doi+10.1007/JHEP06(2014)097}
  {\textsc{inSPIRE}}].
  
\bibitem{pinkbook}
  R.~K.~Ellis, W.~J.~Stirling and B.~R.~Webber,
  {\em QCD and Collider Physics},
  \href{https://www.cambridge.org/core/books/qcd-and-collider-physics/D0095E6D278BBBC74E9C3636AB4CB80C#}
  {(Cambridge University Press, Cambridge, UK, 1996)} 
  [\href{http://inspirehep.net/record/328604}
  {\textsc{inSPIRE}}].

\bibitem{blackbook} 
  J.~Campbell, J.~Huston and F.~Krauss,
  {\em The Black Book of Quantum Chromodynamics : A Primer for the LHC Era},
  \href{https://global.oup.com/academic/product/the-black-book-of-quantum-chromodynamics-9780199652747}
  {(Oxford University Press, Oxford, UK, 2018)} 
  [\href{http://inspirehep.net/record/1635686}
  {\textsc{inSPIRE}}].

\bibitem{CT14}
  S.~Dulat {\it et al.},
  {\em New parton distribution functions from a global analysis of 
  quantum chromodynamics},
  \href{http://dx.doi.org/10.1103/PhysRevD.93.033006}
  {Phys.\ Rev.\ D {\bf 93}, 033006 (2016)}
  [\href{http://inspirehep.net/search?p=doi:10.1103/PhysRevD.93.033006}
  {\textsc{inSPIRE}}].

\bibitem{NSThresholdII}
  Z.~Nagy and D.~E.~Soper,
  {\em Jets and threshold summation in Deductor},
  \href{http://dx.doi.org/10.1103/PhysRevD.98.014035}
  {Phys.\ Rev.\ D {\bf 98}, 014035 (2018)}
  [\href{http://inspirehep.net/search?p=find+doi+10.1103/PhysRevD.98.014035}
  {\textsc{inSPIRE}}].

\bibitem{NSAllOrder} 
  Z.~Nagy and D.~E.~Soper,
  {\em What is a parton shower?},
  \href{http://dx.doi.org/10.1103/PhysRevD.98.014034}
  {Phys.\ Rev.\ D {\bf 98}, 014034 (2018)}
  [\href{http://inspirehep.net/search?p=find+doi+10.1103/PhysRevD.98.014034}
  {\textsc{inSPIRE}}].


\bibitem{NSI}
  Z.~Nagy and D.~E.~Soper,
  {\em Parton showers with quantum interference},
  \href{http://dx.doi.org/10.1088/1126-6708/2007/09/114}
  {JHEP {\bf 0709}, 114 (2007)}
  [\href{http://inspirehep.net/search?p=doi:10.1088/1126-6708/2007/09/114}
  {\textsc{inSPIRE}}].

\bibitem{NSII}
  Z.~Nagy and D.~E.~Soper,
  {\em Parton showers with quantum interference: Leading color, spin averaged},
  \href{http://dx.doi.org/10.1088/1126-6708/2008/03/030}
  {JHEP {\bf 0803}, 030 (2008)}
  [\href{http://inspirehep.net/search?p=doi:10.1088/1126-6708/2008/03/030}
  {\textsc{inSPIRE}}].

\bibitem{NSspin}
  Z.~Nagy and D.~E.~Soper,
  {\em Parton showers with quantum interference: Leading color, with spin},
  \href{http://dx.doi.org/10.1088/1126-6708/2008/07/025}
  {JHEP {\bf 0807}, 025 (2008)}
  [\href{http://inspirehep.net/search?p=find+doi+10.1088/1126-6708/2008/07/025}
  {\textsc{inSPIRE}}].
  
\bibitem{NScolor}
  Z.~Nagy and D.~E.~Soper,
  {\em Parton shower evolution with subleading color},
  \href{http://dx.doi.org/10.1007/JHEP06(2012)044}
  {JHEP {\bf 1206}, 044 (2012)}
  [\href{http://inspirehep.net/search?p=find+doi+10.1007/JHEP06(2012)044}
  {\textsc{inSPIRE}}].

\bibitem{ShowerTime}
  Z.~Nagy and D.~E.~Soper,
  {\em Ordering variable for parton showers},
  \href{http://dx.doi.org/10.1007/JHEP06(2014)178}
  {JHEP {\bf 1406}, 178 (2014)}
  [\href{http://inspirehep.net/search?p=find+doi+10.1007/JHEP06(2014)178}
  {\textsc{inSPIRE}}].

\bibitem{PartonDistFctns}
  Z.~Nagy and D.~E.~Soper,
  {\em Parton distribution functions in the context of parton showers},
  \href{http://dx.doi.org/10.1007/JHEP06(2014)179}
  {JHEP {\bf 1406}, 179 (2014)}
  [\href{http://inspirehep.net/search?p=find+doi+10.1007/JHEP06(2014)179}
  {\textsc{inSPIRE}}].

\bibitem{NSThreshold}
  Z.~Nagy and D.~E.~Soper,
  {\em Summing threshold logs in a parton shower},
  \href{http://dx.doi.org/10.1007/JHEP10(2016)019}
  {JHEP {\bf 1610}, 019 (2016)}
  [\href{http://inspirehep.net/search?p=find+doi+10.1007/JHEP10(2016)019}
  {\textsc{inSPIRE}}].

\bibitem{NSbettercolor} 
  Z.~Nagy and D.~E.~Soper,
  {\em Parton showers with more exact color evolution},
  \href{http://dx.doi.org/10.1103/PhysRevD.99.054009}
  {Phys.\ Rev.\ D {\bf 99}, 054009 (2019)}
  [\href{http://inspirehep.net/search?p=find+doi+10.1103/PhysRevD.99.054009}
  {\textsc{inSPIRE}}].

\bibitem{NSexpipi} 
  Z.~Nagy and D.~E.~Soper,
  {\em Exponentiating virtual imaginary contributions in a parton shower},
  \href{http://dx.doi.org/10.1103/PhysRevD.100.074005}
  {Phys.\ Rev.\ D {\bf 100}, 074005 (2019)}
  [\href{http://inspirehep.net/search?p=find+doi+10.1103/PhysRevD.100.074005}
  {\textsc{inSPIRE}}].



\bibitem{Sterman1987}
  G.~F.~Sterman,
  {\em Summation of Large Corrections to Short Distance Hadronic 
  Cross-Sections},
  \href{http://dx.doi.org/10.1016/0550-3213(87)90258-6}
  {Nucl.\ Phys.\ B {\bf 281}, 310 (1987)}\
  [\href{http://inspirehep.net/search?p=find+doi+10.1016/0550-3213(87)90258-6}
  {\textsc{inSPIRE}}].
  
\bibitem{Catani32}
  S.~Catani and L.~Trentadue,
  {\em Resummation of the QCD Perturbative Series for Hard Processes},
  \href{http://dx.doi.org/10.1016/0550-3213(89)90273-3}
  {Nucl.\ Phys.\ B {\bf 327}, 323 (1989)}
  [\href{http://inspirehep.net/search?p=find+doi+10.1016/0550-3213(89)90273-3}
  {\textsc{inSPIRE}}].

\bibitem{CMW}
  S.~Catani, B.~R.~Webber and G.~Marchesini,
  {\em QCD coherent branching and semiinclusive processes at large x},
  \href{http://dx.doi.org/10.1016/0550-3213(91)90390-J}
  {Nucl.\ Phys.\ B {\bf 349} (1991) 635}
  [\href{http://inspirehep.net/search?p=find+doi+10.1016/0550-3213(91)90390-J}
  {\textsc{inSPIRE}}].

\bibitem{CataniManganoNason32}
  S.~Catani, M.~L.~Mangano, P.~Nason and L.~Trentadue,
  {\em The Resummation of soft gluons in hadronic collisions},
  \href{http://dx.doi.org/10.1016/0550-3213(96)00399-9}
  {Nucl.\ Phys.\ B {\bf 478}, 273 (1996)}
  [\href{http://inspirehep.net/search?p=find+doi+10.1016/0550-3213(96)00399-9}
  {\textsc{inSPIRE}}]
  
\bibitem{SudakovFactorization}
  H.~Contopanagos, E.~Laenen and G.~F.~Sterman,
  {\em Sudakov factorization and resummation},
  \href{http://dx.doi.org/10.1016/S0550-3213(96)00567-6}
  {Nucl.\ Phys.\ B {\bf 484}, 303 (1997)}
  [\href{http://inspirehep.net/search?p=find+doi+10.1016/S0550-3213(96)00567-6}
  {\textsc{inSPIRE}}].
  
\bibitem{ManoharSCET}
  A.~V.~Manohar,
  {\em Deep inelastic scattering as $x \to 1$ using 
  soft collinear effective theory},
  \href{http://dx.doi.org/10.1103/PhysRevD.68.114019}
  {Phys.\ Rev.\ D {\bf 68}, 114019 (2003)}
  [\href{http://inspirehep.net/search?p=find+doi+10.1103/PhysRevD.68.114019}
  {\textsc{inSPIRE}}].

  \bibitem{BecherNeubertPecjak}
  T.~Becher, M.~Neubert and B.~D.~Pecjak,
  {\em Factorization and Momentum-Space Resummation in 
  Deep-Inelastic Scattering},
  \href{http://dx.doi.org/10.1088/1126-6708/2007/01/076}
  {JHEP {\bf 0701}, 076 (2007)}
  [\href{http://inspirehep.net/search?p=find+doi+10.1088/1126-6708/2007/01/076}
  {\textsc{inSPIRE}}].

\bibitem{Stewart:2009yx}
  I.~W.~Stewart, F.~J.~Tackmann and W.~J.~Waalewijn,
  {\em Factorization at the LHC: From PDFs to Initial State Jets},
  \href{http://dx.doi.org/10.1103/PhysRevD.81.094035}
  {Phys.\ Rev.\ D {\bf 81}, 094035 (2010)}
  [\href{http://inspirehep.net/search?p=find+doi+10.1103/PhysRevD.81.094035}
  {\textsc{inSPIRE}}].
  

\end{thebibliography}
\end{document}